\documentclass[longbibliography,preprint,aps,prX,citeautoscript,superscriptaddress,11pt]{revtex4-1}

\usepackage[utf8]{inputenc}
\usepackage{graphicx}
\usepackage{amsmath}
\usepackage{hyperref}
\usepackage{textcomp}
\usepackage{newtxmath}

\begin{document}

\title{Skyrmion Topological Hall Effect near Room Temperature}

\author{Maxime Leroux}
\affiliation{Materials Physics and Applications Division, Los Alamos National Laboratory, Los Alamos, New Mexico 87545, United States}

\author{Matthew J. Stolt}
\affiliation{Department of Chemistry, University of Wisconsin-Madison, 1101 University Avenue, Madison, Wisconsin 53706, USA}

\author{Song Jin}
\affiliation{Department of Chemistry, University of Wisconsin-Madison, 1101 University Avenue, Madison, Wisconsin 53706, USA}

\author{Douglas V. Pete}
\affiliation{Center for Integrated Nanotechnologies, Sandia National Laboratories, Albuquerque, New Mexico 87185, United States}

\author{Charles Reichhardt}
\affiliation{Theoretical Division, Los Alamos National Laboratory, Los Alamos, New Mexico 87545, United States}

\author{Boris Maiorov}
\thanks{Corresponding author: \href{mailto:maiorov@lanl.gov}{maiorov@lanl.gov}}
\affiliation{Materials Physics and Applications Division, Los Alamos National Laboratory, Los Alamos, New Mexico 87545, United States}

\begin{abstract}
Magnetic skyrmions are stable nanosized spin structures that can be displaced at low electrical current densities. Because of these properties, they have been proposed as building blocks of future electronic devices with unprecedentedly high information density and low energy consumption. The electrical detection of skyrmions via the Topological Hall Effect (THE), has so far been demonstrated only at cryogenic temperatures. Here, we report the observation of a skyrmion Topological Hall Effect near room temperature (276\,K) in a mesoscopic lamella of FeGe. This region unambiguously coincides with the skyrmion lattice location revealed by neutron scattering. We provide clear evidence of a reentrant helicoid magnetic phase adjacent to the skyrmion phase, and discuss the large THE amplitude (5\,n$\Omega$.cm) in view of the ordinary Hall Effect.
\end{abstract}

\maketitle


Magnetic skyrmions are nanometre size quasiparticles with a whirling spin configuration, which hold great potential as information carriers for electronic devices\cite{Nagaosa2013NatNanoReview,Fert_Reyren_Cros_2017NatRevMat,Fert2013,Finocchio2016JoPDReview,Liu2015ChinPhysBReview}. Magnetic skyrmions were first studied theoretically\cite{belavin1975skyrmions,bogdanov1989skyrmions,Roessler2006}, and have recently been evidenced experimentally in B20 chiral materials\cite{Muhlbauer2009ScienceMnSineutron,Yu2011NatMat} and in multilayer ferromagnetic thin films\cite{Boulle2016NatNano}.
Most skyrmionic materials have in common\cite{Nagaosa2013NatNanoReview,Fert_Reyren_Cros_2017NatRevMat} the presence of an antisymmetric exchange interaction, known as Dzyaloshinskii–Moriya (DM) interaction\cite{dzialoshinskii1957,Moriya1960}, favoring spin canting in materials that would otherwise be ferromagnetic with parallel aligned spins. 
The DM interaction occurs in materials with spin–orbit coupling and a structure lacking inversion symmetry. In B20 magnets such as MnSi\cite{Muhlbauer2009ScienceMnSineutron} and FeGe\cite{Yu2011NatMat}, this emerges from a heavy magnetic element in a non-centrosymmetric crystalline structure. In multilayer ferromagnetic thin films, the DM interaction is engineered by combining ultrathin layers of heavy metal and ferromagnetic materials, in which the heavy metal provides the large spin-orbit coupling while the interface between layers breaks the inversion symmetry\cite{Fert2013}.
For a large enough DM interaction, such as that found in B20 magnets, the ground state of the material is helical. In this state, the spins form helices with a pitch of a few tens of nanometers typically. This pitch scales as J/D, where D is the DM interaction constant and J is the exchange energy\cite{Nagaosa2013NatNanoReview}. This pitch is referred to as the helical wavelength and it sets the skyrmion size in bulk materials.

The topologically non-trivial whirling configuration of skyrmions spin texture can be characterized by the topological charge  that is $\pm1$ for an individual skyrmion, and null for topologically trivial structures such as magnetic bubbles. The topologically non-trivial structure of magnetic skyrmions makes them relatively stable because they cannot be continuously deformed to another magnetic state\cite{Nagaosa2013NatNanoReview,Braun2012}.

A key advantage of magnetic skyrmions as stable nano-objects is that they can be displaced using relatively low electrical currents\cite{Jonietz2011Science,Schulz2012NatPhysMnSiJc,Fert2013}. This low critical current stems from the relative insensitivity of skyrmions to disorder due to the strong Magnus force contribution to the skyrmion dynamics that allows skyrmions to bypass defects\cite{Finocchio2016JoPDReview,Liu2015ChinPhysBReview,Nagaosa2013NatNanoReview,reichhardt1}. 
Such current manipulation enables not only data storage but also logic devices using magnetic ``racetrack'' type circuits\cite{Fert_Reyren_Cros_2017NatRevMat,Fert2013}. Skyrmions nanometer size (10\,nm = 6\,Tbit/in$^2$), stability and low critical current (10$^6$\, A/m$^2$) could thus yield non-volatile electronic devices with unprecedentedly high information density and low energy consumption.
%

One of the key issues for future ``skyrmionic'' devices is to develop a method for detecting skyrmions through electrical means. Electrical manipulation of skyrmions was demonstrated by visualization techniques at room temperature a few years ago\cite{iwasaki2013current,zhang2015magnetic,jiang2015blowing,buttner2015dynamics,Woo2016,Juge2017a}. Yet it is only very recently, probably because of the sensitive measurements required, that electrical detection of skyrmions was demonstrated for isolated metastable skyrmions generated by interfacial DM interaction in multilayer thin films\cite{Maccariello2018a} using the Anomalous Hall Effect (AHE). This detection relied on the large change of magnetization, and the concomitant large change in anomalous Hall Effect, introduced by a skyrmion in a ferromagnetic background. This effect is however not readily applicable to bulk skyrmion compounds, so-called B20 chiral magnets, because the skyrmion lattice does not introduce a significant step-change in magnetization over the background. Indeed, because of the larger DM interaction in B20 compounds, the background that surrounds the thermodynamically stable skyrmion lattice phase is a conical phase with a winding spins configuration, rather than a ferromagnetic phase.

Thus, a more practical way of detecting skyrmions in bulk compounds is to use skyrmions Topological Hall Effect (THE). This effect originates from electrons accumulating a Berry phase as they travel through skyrmions spin configuration, which acts as an effective magnetic field ($B_{eff}$). To first order, the skyrmion THE amplitude is proportional to $B_{eff}$, which itself is inversely proportional to a skyrmion cross-section area. The skyrmion THE is usually a few n$\Omega$.cm, and it arises in addition to the ordinary (OHE) and anomalous Hall Effect (AHE) in the transverse resistivity of skyrmionic compounds.
The skyrmion THE has so far only been detected and unambiguously correlated with skyrmions imaging in the MnSi family of compounds at cryogenic temperatures\cite{Neubauer2009PRLTHE_MnSi,Schulz2012NatPhysMnSiJc,Franz2014PRL_HallMnCoFeSi,dong2015}.

Recently, two bulk materials have been unambiguously shown by Lorentz Transmission Electron Microscopy (LTEM) and Small Angle Neutron Scattering (SANS) to have a skyrmions lattice near or at room temperature: Co-Zn-Mn alloys\cite{Tokunaga2015NatComCoZnMn,Yu2017advMat}, and cubic FeGe\cite{Yu2011NatMat,Yu2012NatComFeGe_Jc,McGrouther2016NJpP_TEM_FeGe,
StoltJin2016nanolet_FeGeNW,Moskvin2013PRLSANSFeGe}.
In FeGe, magnetization\cite{Wilhelm2011PRLFeGemultiphase}, specific heat measurements\cite{CeveyPSS2013FeGe}, and microwave absorption spectroscopy\cite{StoltJin2017} also identified several magnetic phases compatible with skyrmions. 
But bulk skyrmions THE near room temperature had so far not been observed.
A few transport studies in FeGe thin films\cite{Huang2012PRL_FeGe_rho_film,Porter2014PRBFeGefilmTransport,Kanazawa2015PRB_FeGeconstricted} did report very large Hall signature at cryogenic temperatures. However, skyrmions have not been confirmed by imaging techniques in thin films of FeGe\cite{Zhang2017SciRepFeGefilmswithoutSkX}. All LTEM measurements were performed on lamellae extracted from bulk FeGe single crystals \cite{Yu2011NatMat,Yu2012NatComFeGe_Jc,McGrouther2016NJpP_TEM_FeGe}. Thus, it is therefore still debated whether the unusually large Hall signal, orders of magnitude larger than in bulk MnSi, observed in thin films at cryogenic temperature comes from a skyrmionic THE or has another origin\cite{MoncheskyPRBquestionFeGethinfilms,MoncheskyCommenttoPRL}.

Here, we demonstrate the first electrical detection of skyrmions near room temperature in a bulk crystal: using high sensitivity Hall Effect measurements in FeGe, we observe a skyrmion THE of amplitude +5 n$\Omega$.cm at 276\,K (3°C). This THE unambiguously coincides with the skyrmion lattice phase identified by small angle neutron scattering\cite{Moskvin2013PRLSANSFeGe}, and its amplitude is as large as the THE observed in MnSi at 29\,K. Adjacent to the skyrmion phase, we also observe the signature of a magnetic phase that is not present in the prototypical skyrmion compound MnSi, and merges into the inhomogeneous chiral spin state\cite{Wilhelm2016PRB_FeGescaling}. Finally, comparing THE in MnSi and FeGe we emphasize the failings of the formula that is generally used to estimate the skyrmion THE.

\section*{Results and Discussion}

\begin{figure}
  \includegraphics[width=0.5\columnwidth]{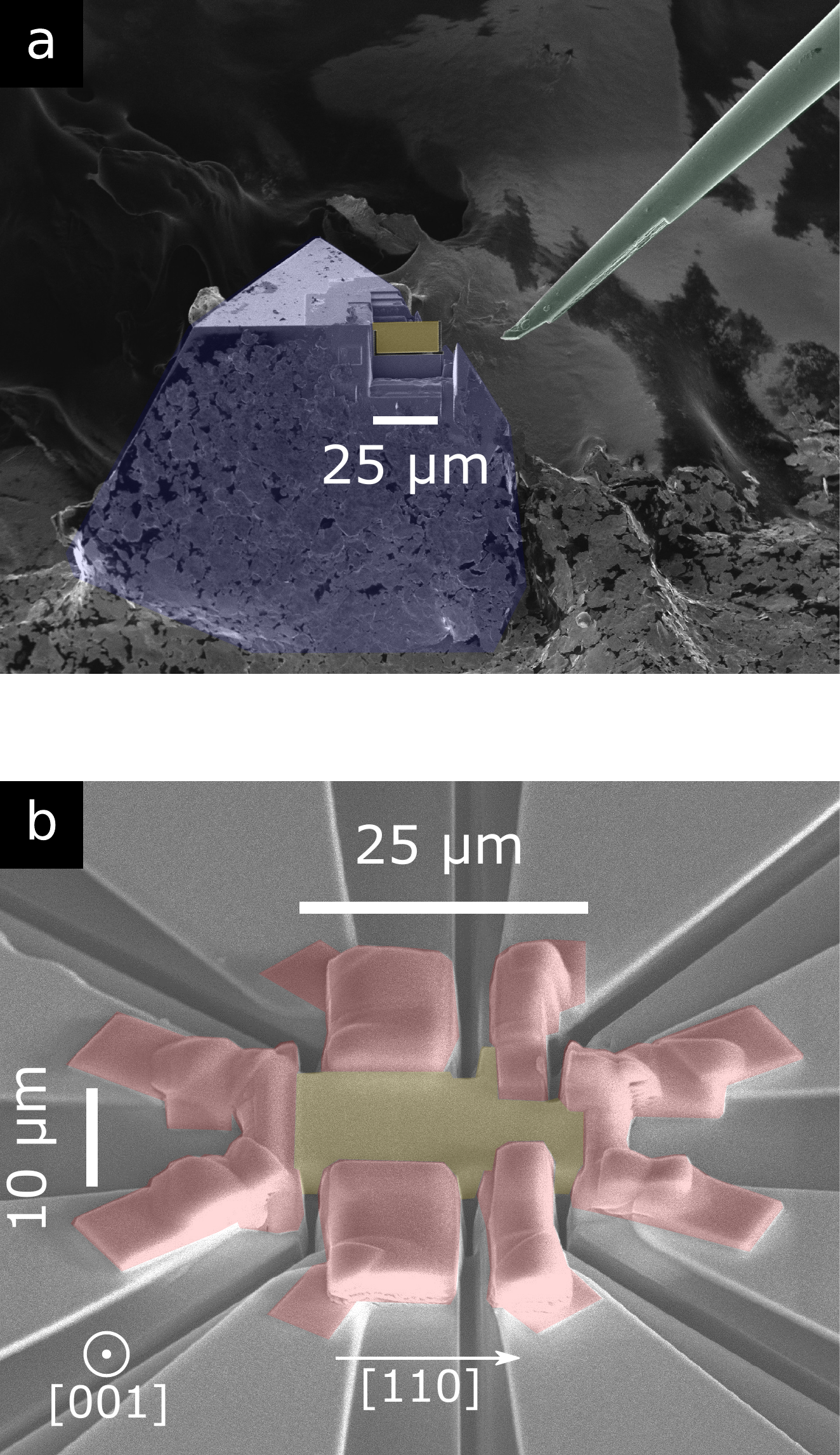}
  \caption{\textbf{Nanofabrication of electrical transport sample by Focused Ion Beam (FIB).} \textbf{(a)} Electron microscope picture. False colors: \textit{(yellow)} lamella sample, \textit{(blue)} FeGe single crystal, \textit{(green)} FIB nanoprobe tip. FeGe lamella sample carved from a pyramidal FeGe single crystal, before final cut and lift-out with nanoprobe. The single crystal is electrically grounded to the FIB sample holder with silver paint. \textbf{(b)} Lamella with platinum electrodes on silicon oxide chip. Final dimensions: $10\times25\times0.75$\,\textmu{}m$^3$. The electrical current flows along [110] (see arrow) and the magnetic field is applied perpendicular to the flat face of the lamella along [001]. False colors: \textit{(yellow)} lamella sample, \textit{(red)} platinum contacts deposited by FIB, \textit{(light gray)} evaporated platinum strips, \textit{(dark gray)} silicon oxide chip.  
}
  \label{figA}
\end{figure}

\subsection*{Electrical Transport Measurements}
\begin{figure}
  \includegraphics[width=0.5\columnwidth]{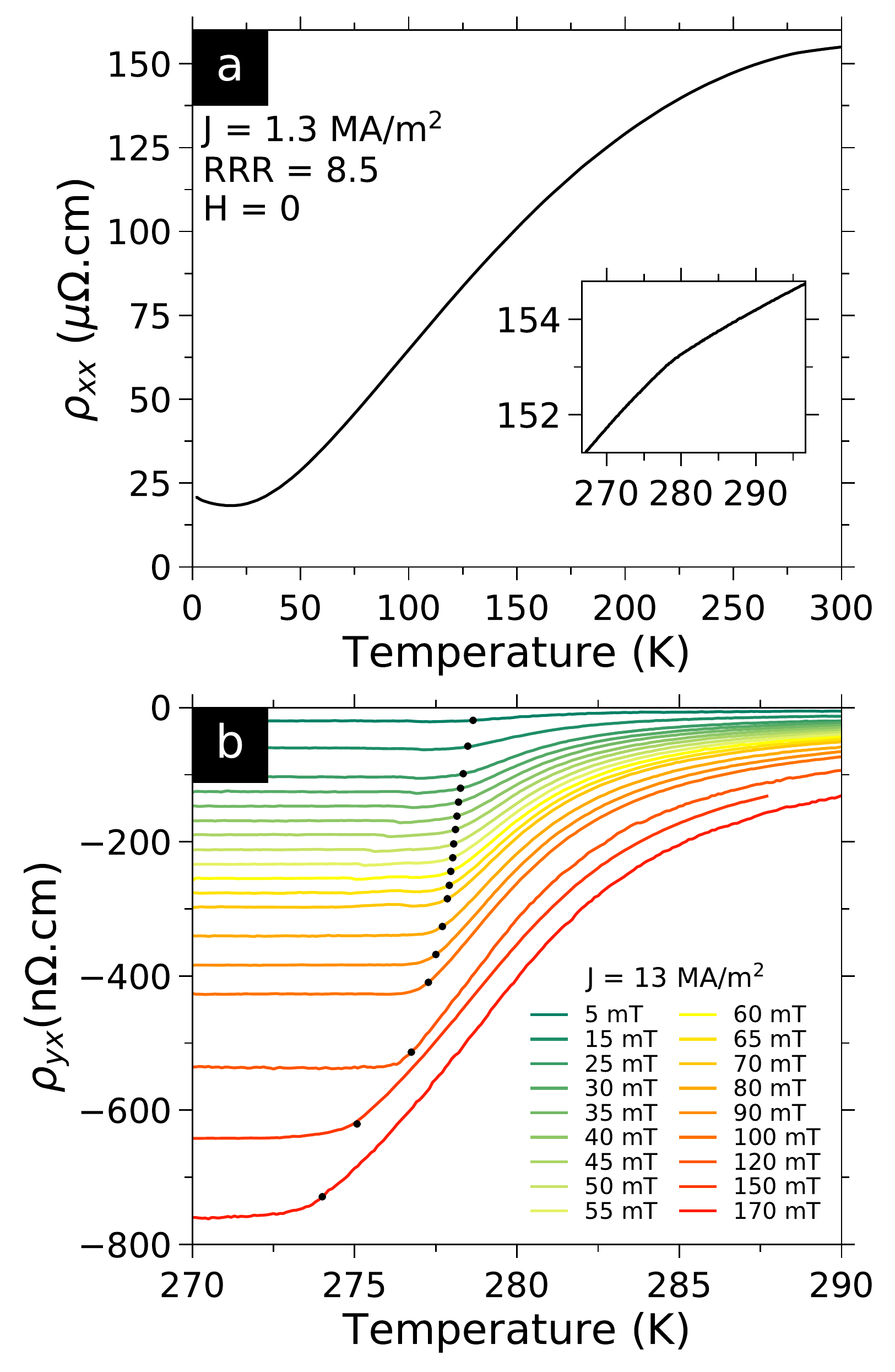}
  \caption{\textbf{Electrical transport properties of the FeGe lamella.}  \textbf{(a)} Temperature dependence of the zero field longitudinal resistivity $\rho_{xx}$ of the sample, in-line with the literature. \textbf{(Inset)} Zoom on the kink around 279\,K marking the magnetic transition. \textbf{(b)} Temperature dependence of the transverse resistivity $\rho_{yx}$ anti-symmetrized with respect to the magnetic field. The change in $\rho_{yx}$ marks the magnetic transition between the conical phase at low temperature and the paramagnetic/field polarized phases at high temperature. Black circles indicate the transition temperature defined using a 5\% criterion and reported in Fig.~\ref{figD}.}
  \label{figB}
\end{figure}

We prepared a 0.75\,\textmu{}m thick lamella of FeGe from a single crystal using a Focused Ion Beam microscope (FIB). Fig.~\ref{figA}.a shows the lamella before lift-out from the single crystal, and Fig.~\ref{figA}.b shows the lamella with Pt electrodes deposited by FIB. We then measured the electrical transport properties of this lamella which shows transport properties consistent with bulk crystals (see SI). Fig.~\ref{figB}.a. shows the temperature dependence of the longitudinal resistivity $\rho_{xx}$ in zero magnetic field. The overall dependence and absolute value of $\rho_{xx}$ is in agreement with the literature\cite{Porter2014PRBFeGefilmTransport}. The residual resistivity ratio is 8.5, higher than the value of 4\,--\,5 typically found in thin films\cite{Porter2014PRBFeGefilmTransport,Huang2012PRL_FeGe_rho_film} but below the values of 13 to 25 found in large single crystals\cite{Yeo2003PRLFeSiGe,Pedrazzini2007PRLFeGe_rhoxx_vs_P}. The magnetic transition from the paramagnetic phase to the helical phase appears as a subtle but clear kink around 279\,K in the zero-field $\rho_{xx}$ data (inset of Fig.~\ref{figB}.a).

In Fig.~\ref{figB}.b., we show the temperature dependence of the Hall Effect $\rho_{yx}$ of the lamella (using the same sign convention as in Ref.~\cite{Neubauer2009PRL_THE_MnSi,Huang2012PRL_FeGe_rho_film,Porter2014PRBFeGefilmTransport}, that $\rho_{yx}$ is positive for electrons dominated transport). Electrical conduction appears to be predominantly hole type as $\rho_{yx}$ is negative around $T_N$ (270-290\,K). The onset of the increase in $\rho_{yx}$ with increasing temperature, marks the transition from the conical to the paramagnetic phase indicated by black circles in Fig.\ref{figB}. All $\rho_{yx}(T)$ curves are constant in the conical phase in the temperature range we measured. For each magnetic field value, we define a transition temperature as 5\% of the increase in $\rho_{yx}(T)$ between the flat baseline at 270\,K and the value at 290\,K. The magnetic field dependence of this transition temperature is in good agreement with previous magnetization studies in a bulk sample\cite{Wilhelm2011PRLFeGemultiphase,Wilhelm2016PRB_FeGescaling}.

\begin{figure}
  \includegraphics[width=0.8\columnwidth]{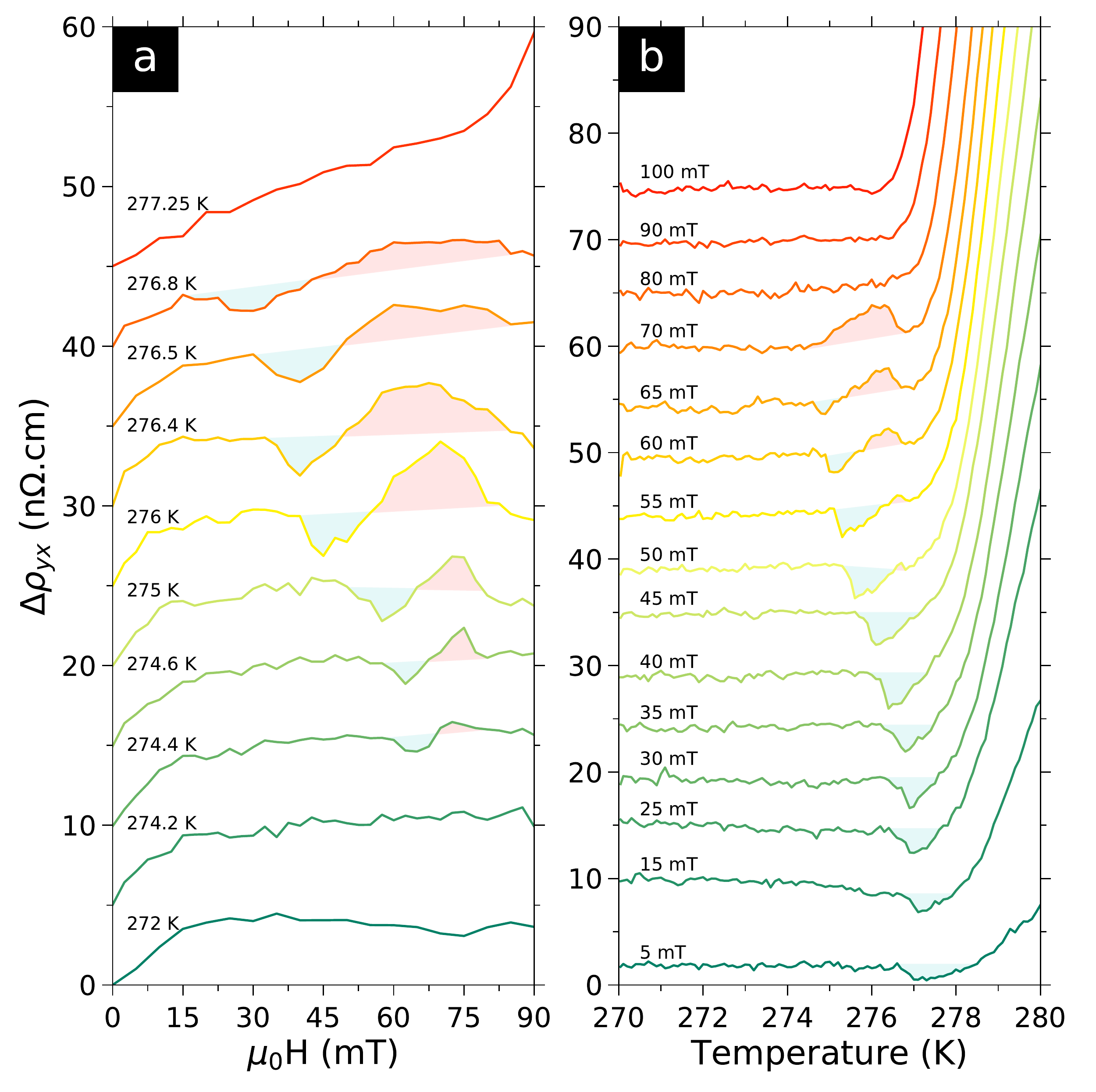}
 \caption{\textbf{Temperature and field dependence of the deviation to the linear normal and anomalous Hall Effects.} \textbf{(a)} Deviation $\Delta\rho_{yx}$ (see text) as a function of applied magnetic field for several temperatures. Curves were shifted up for clarity. The helical phase occurs below the slope change at $\approx15\,$mT, while the conical phase occurs above. The field polarized state appears as a rapid upturn at higher field (shown at 277.25\,K). The blue shaded regions emphasize the local minimum which appears to emerge from the zero field magnetic transition (cf. Fig.~\ref{figD}). The origin of this minimum is still unidentified but might be a reentrant ICS state. The red shaded regions emphasize the local maximum which coincides with the skyrmions lattice phase identified by SANS\cite{Moskvin2013PRLSANSFeGe} in bulk crystals (cf. Fig.~\ref{figD}). We attribute this maximum to the THE of skyrmions. \textbf{(b)} Deviation $\Delta\rho_{yx}$ as a function of temperature for several applied magnetic fields. Curves were shifted up for clarity. The same features are observed.}
  \label{figC}
\end{figure}

The Hall Effect in skyrmionic compounds typically consists of three terms
\begin{equation}
\label{eq:rhoyxdecomp}
\rho_{yx} = \rho^\mathrm{OHE} + \rho^\mathrm{AHE} +\rho^\mathrm{THE}
\end{equation}
The first term is the Ordinary Hall Effect $\rho^\mathrm{OHE}=R_0\mu_0H$ proportional to the applied magnetic field, the second term is the Anomalous Hall Effect $\rho^\mathrm{AHE}=R_sM$ proportional to the magnetization $M$ of the sample and the third term is the THE induced by the spin texture of skyrmions. Here, we find that $\rho_{yx}(H)$ curves are linear in the conical phase, as expected. Indeed the Ordinary Hall Effect and the Anomalous Hall Effect are both linear in H, as the magnetization is proportional to H in this phase\cite{Wilhelm2011PRLFeGemultiphase}. 
So, to better see variations from the conical state, we subtracted the total linear contribution from $\rho_{yx}(H)$ curves in the same manner as Ref.~\cite{Neubauer2009PRL_THE_MnSi}.
The slope of the linear fit of $\rho_{yx}(H)$ in the conical phase (between 270 and 271K), is $-4.3164\,$n$\Omega$.cm/mT with $r^2 = 0.99998$. We subtracted this slope from all $\rho_{yx}(H)$ curves and defined the resulting deviation of $\rho_{yx}(H)$ to the conical phase as $\Delta\rho_{yx}$. $\Delta\rho_{yx}(H)$ curves are shown in Fig.~\ref{figC}.a (shifted for clarity). 
As the conical contribution is constant in temperature, converting $\rho_{yx}(T)$ curves to $\Delta\rho_{yx}(T)$ corresponds to adding an offset to each curve shown in Fig.~\ref{figB}.b. 

\subsection*{Signatures in the Hall Effect}

In Fig.~\ref{figC}.a, the $\Delta\rho_{yx}(H)$ curves below 274.4\,K show several features. A change in slope at $\approx15\,$mT separates the helical and conical phases, at low and high field respectively. $\Delta\rho_{yx}(H)$ curves are flat in the conical phase, consistent with the definition of $\Delta\rho_{yx}$. The upturn at higher fields (shown at 277.25\,K) corresponds to the field polarized state.
The helical-to-conical signature disappears above $\approx277\,$K, in agreement with magnetization studies\cite{Wilhelm2011PRLFeGemultiphase,Wilhelm2016PRB_FeGescaling}.
In addition to the helical, conical and field polarized states, we observe a local maximum and minimum that appear simultaneously inside the conical phase. The maximum and minimum have maximum amplitude at 276\,K of $\approx +5\,$n$\Omega$.cm and $\approx -3\,$n$\Omega$.cm, respectively.

\begin{figure}
  \includegraphics[width=0.65\columnwidth]{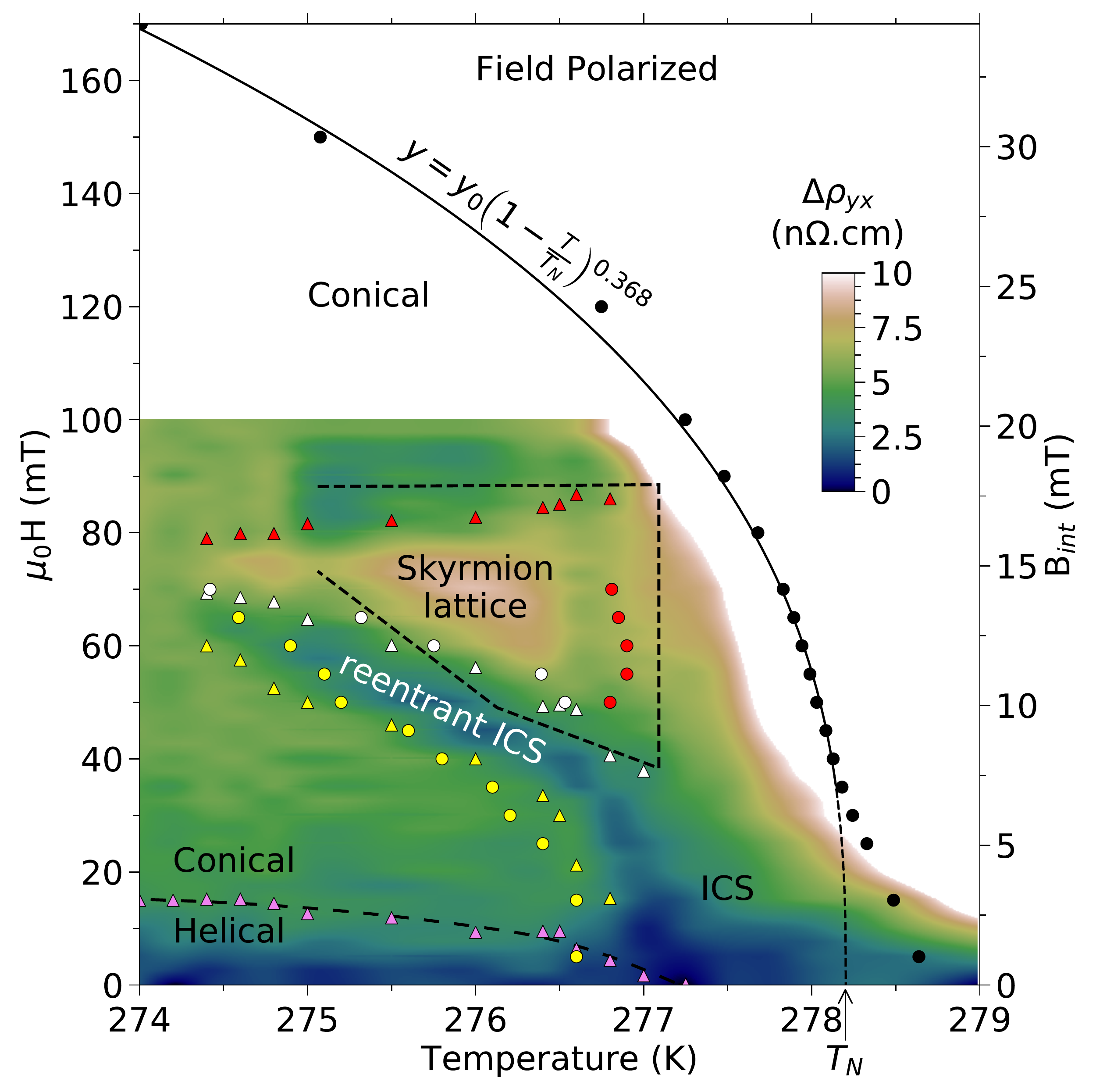}
  \caption{\textbf{HT diagram of the deviation to the linear Hall Effect.} Deviation $\Delta\rho_{yx}(T,H)$  (see text) deduced from $\rho_{yx}(H)$ and $\rho_{yx}(T)$ curves. \textit{(Black circles)} transition separating the conical and field polarized phases, as defined by the 5\% criterion in $\rho_{yx}(T)$ curves in Fig.~\ref{figB}.b. The solid line is a fit to the high field part with critical exponent $\beta=0.368$ for 3D Heisenberg spins, as previously observed\cite{Wilhelm2016PRB_FeGescaling}. T$_N$ extrapolates to $278.2\,$K, also in-line with the literature\cite{Wilhelm2016PRB_FeGescaling}. \textit{(Pink triangles)} low field change in slope in $\Delta\rho_{yx}(H)$ curves, coinciding with the helical to conical transition. Dashed line is a guide to the eye. \textit{(Yellow triangles and circles)} left onset of the local minimum in $\Delta\rho_{yx}(H)$ and $\Delta\rho_{yx}(T)$, respectively. \textit{(White triangles and circles)} point of inversion between the minimum and maximum. \textit{(Red triangles and circles)} right onset of the maximum. \textit{(Dotted edge polygon)} Skyrmion lattice phase measured by SANS\cite{Moskvin2013PRLSANSFeGe} for H//[100] in a spherical bulk crystal after correcting for demagnetizing effects. From this, we attribute the maximum in $\Delta\rho_{yx}$ to the topological Hall Effect of the skyrmion lattice. No SANS data in the longitudinal geometry showing the six-fold scattering, is published at temperature below the open end of the polygon. The origin of the local minimum is still unidentified but it appears to continue into the inhomogeneous chiral spin state\cite{Wilhelm2016PRB_FeGescaling} between $T_N$ and the helical state.
  }
  \label{figD}
\end{figure}

In Fig.~\ref{figC}.b, from 270 to $\approx 274.5-277$\,K, $\Delta\rho_{yx}(T)$ is constant in temperature. This corresponds to the helical or conical states, at low or high magnetic fields respectively. Above $\approx278$\,K there is a large upturn corresponding to the transition to the paramagnetic or field polarized states.
Around 276\,K, we also observe a local maximum and minimum. The minimum is present from 5 to 65\,mT, whereas the maximum is present in a narrower field range from 55 to 70\,mT. The maximum and minimum observed in $\Delta\rho_{yx}(H)$ and $\Delta\rho_{yx}(T)$ are consistent in both locations and amplitudes, and have never been reported before. To gain further insights on their origin we combined $\Delta\rho_{yx}(H)$ and $\Delta\rho_{yx}(T)$ data sets in Fig.~\ref{figD} to produce a color map (color maps made from either data sets are shown in the SI).

\subsection*{Skyrmion Topological Hall Effect}

Fig.~\ref{figD} shows that the local maximum occurs in a region of field and temperature, with a shape and location reminiscent of the skyrmion phase in MnSi.
After accounting for demagnetizing effects (right axis of the H-T diagram) following the process of Ref.~\cite{Ritz2010PhDThesis} (see SI), this region corresponds quantitatively in field and temperature to the region where a skyrmion lattice has been observed in bulk crystals of FeGe by SANS\cite{Moskvin2013PRLSANSFeGe}. 
We therefore conclude that this local maximum is the Topological Hall Effect induced by skyrmions magnetic texture. We note that this is the first report of a skyrmion Topological Hall Effect near room temperature.

The maximum amplitude of this THE is $\approx+5$\,n$\Omega$.cm at 276\,K, a value very similar to that of MnSi\cite{Neubauer2009PRL_THE_MnSi} at 29\,K. This is contrary to what is expected from the small $B_{eff}$ of FeGe produced by its larger skyrmions. However, $B_{eff}$ is not the only factor determining the THE amplitude. The exact amplitude of the THE depends on the details of the band structure, but the THE is usually approximated as\cite{Ritz2013PRB_THE_pressure,Neubauer2009PRL_THE_MnSi,EverschorBerryphaseTutorial}:
\begin{equation}
\Delta \rho_{yx}^{max} \approx P.B_{eff}.R_0
\label{THEapprox}
\end{equation}
where $P$ is the local spin-polarization of the conduction electrons in the skyrmion state, $B_{eff}$ is the effective magnetic field and $R_0$ is the ordinary Hall Effect constant. Here, we find (see methods) that in FeGe:
\begin{equation}
\Delta \rho_{yx}^{THE} = 0.082\times-0.74\times-0.73\ 10^{-8} \approx +45\,\mathrm{n}\Omega\mathrm{.cm}
\end{equation}
And this value is of the same order as the approximate THE value in MnSi:
\begin{equation}
\Delta \rho_{yx}^{THE} = 0.09\times-13.15\times0.12\ 10^{-8} \approx -140\,\mathrm{n}\Omega\mathrm{.cm}
\end{equation}
We note that P is similar in both compounds, $B_{eff}$ is 18 times smaller in FeGe than in MnSi, and $R_0$ is 6 times larger in FeGe than in MnSi. Overall, the THE amplitude estimated via Eqn.~\ref{THEapprox} is thus only three times smaller in FeGe than in MnSi. 

Both THE values clearly overestimate the experimental values of the THE ($\approx +5\,$n$\Omega$.cm in both MnSi and FeGe). This overestimation has been known for some time in MnSi and attributed to the factor $R_0$\cite{Ritz2013PRB_THE_pressure}. Recently, it was pointed out that this estimate is only an upper bound of skyrmions THE\cite{Maccariello2018a}. We also point out that the sign of the THE calculated using Eqn.~\ref{THEapprox} appears to be opposite to what is measured in MnSi\cite{Neubauer2009PRL_THE_MnSi}. Recent theoretical efforts\cite{Denisov2017} show that a range of sign and amplitude is to be expected for the THE depending on coupling strength and electronic scattering rates. Looking beyond this ongoing debate, we note that at least the THE estimates for MnSi and FeGe are of the same order of magnitude in absolute value: 140\,$\mathrm{n}\Omega\mathrm{.cm}$ for MnSi and 45\,$\mathrm{n}\Omega\mathrm{.cm}$ for FeGe. Thus, this could explain why the experimental values of the THE are so similar in both compounds.

We also emphasize that the larger $R_0$ of FeGe appears to compensate the smaller $B_{eff}$. This means that the charge carrier density, and more generally the band structure, is a tool as important as skyrmions effective magnetic field in achieving measurable THE. For instance, in Pt/Co/Ir multilayer thin films, the skyrmion THE was estimated to be immeasurably small: $\Delta \rho_{yx}^{THE} = 0.0017\,\mathrm{n}\Omega\mathrm{.cm}$ as $R_0=2.10^{-11}\,\Omega$.m.T$^{-1}$\cite{Maccariello2018a}. This value is more than 300 times smaller than the value $|R_0|=7.3.10^{-9}\,\Omega$.m.T$^{-1}$ in FeGe, which suggest that FeGe is a compound of choice for the study of the THE near room temperature.

\subsection*{Signature of a Reentrant Helicoid Magnetic Phase}

The local minimum observed in $\Delta\rho_{yx}$ in Fig.~\ref{figD}, is new to all compounds with clear signatures of vortex-like cylindrical skyrmions. A positive THE has been observed in MnSi\cite{Neubauer2009PRL_THE_MnSi}, and a negative THE was found in compounds such as Mn$_{1-x}$Fe$_x$Si\cite{Franz2014PRL_HallMnCoFeSi}, but a positive and negative THE have never been observed consecutively, up to our knowledge. 

Polycrystalline MnGe does show adjacent maximum and minimum in $\rho_{yx}$\cite{Kanazawa2016MnGe}, but the amplitude of this effect is 50 times larger ($\approx200\,$n$\Omega$.cm) and spans over 10 Tesla and 100 Kelvin. Thus, this is very different from the phenomenon we observe in FeGe. In addition, the accepted interpretation of the behavior of MnGe does not involve \emph{cylindrical} skyrmions but a spin structure periodic in all three dimensions undergoing a topological phase transition through the pair annihilation of hedgehogs and anti-hedgehogs topological spin singularities\cite{Kanazawa2016MnGe}.

In the $\Delta\rho_{yx}(H)$ curves of Fig.~\ref{figC}.a and the H-T diagram of Fig.~\ref{figD}, the local minimum tracks and ends with the lowest temperature of the skyrmions phase. But at higher temperatures it continues into the inhomogeneous chiral spin state\cite{Wilhelm2016PRB_FeGescaling} (ICS). Thus, the local minimum delimits a region which appears to compete with the skyrmion lattice phase but shares the signature of the helical and ICS states.
Theoretical calculations\cite{Wilhelm2011PRLFeGemultiphase} show that the hierarchy of close energy scales yields several possible magnetic phases in the vicinity of $T_N$, including: +$\pi$ skyrmions, -$\pi$ skyrmions, half-skyrmions squares lattice and reentrant helicoid.
The position of the local minimum in the phase diagram, bordering the skyrmion lattice phase, and its sign opposite to the THE of the skyrmion lattice, suggest +$\pi$ skyrmions with a positive effective magnetic field and negative THE; however, SANS does not show an other ordered skyrmion region\cite{Moskvin2013PRLSANSFeGe}.
Thus, this suggests that the local minimum corresponds to a reentrant ICS or helicoid state. Such reentrant phase has never been observed before in skyrmion lattice systems, up to our knowledge. It also underscores the fact that, although magnetic and transport measurements show features\cite{Wilhelm2011PRLFeGemultiphase,Wilhelm2016PRB_FeGescaling} at similar locations, the latter indicate a very different origin for the reentrant phase.

\subsection*{High Current Density Measurements of Hall Signatures}

\begin{figure}
  \includegraphics[width=\columnwidth]{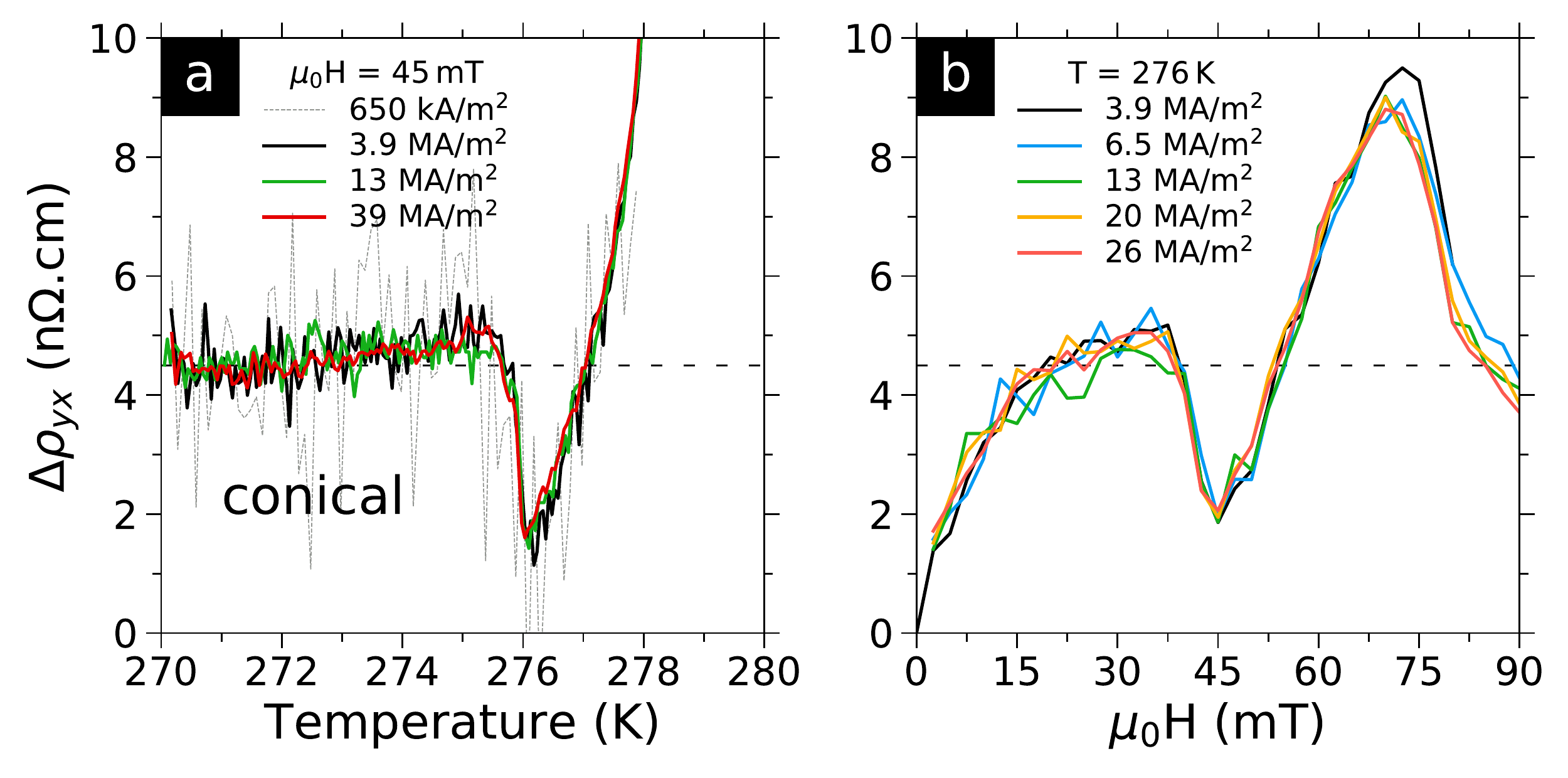}
  \caption{\textbf{Deviation to the linear Hall Effect for several current densities.} \textbf{(a)} $\Delta\rho_{yx}(T)$ curves (see text) at 45\,mT. Only the local minimum is present. No significant changes occur for current densities between 0.65 and 39\,MA/m$^2$, evidencing the absence of self-heating effect. Close inspection of raw data reveals only 10\,mK of heating from 3.9 to 39 MA/m$^2$. \textbf{(b)} $\Delta\rho_{yx}(H)$ curves at 276\,K. No changes occur for current densities between 6.5 and 26\,MA/m$^2$ in either the slope change at $\approx15\,$mT marking the helical to conical phase transition, the local minimum at 40\,mT, and the local maximum at 75\,mT which we attribute to the THE of the skyrmion lattice phase. Data at 3.9\,MA/m$^2$ might show a 0.5\,n$\Omega$.cm increase of the maximum at 75\,mT, which would be compatible with slower or pinned skyrmions, but this is barely above noise level.}
  \label{figE}
\end{figure}

Hall Effect measurements in single crystals of MnSi\cite{Schulz2012NatPhysMnSiJc} show a decrease in THE after skyrmions start to move at a critical current densities of $J_c\approx 0.5$\,MA/m$^2$. Above this threshold, the amplitude of the THE is reduced from $+5\,$n$\Omega$.cm to $+2\,$n$\Omega$.cm as a result of the emerging electric field induced by moving skyrmions\cite{Schulz2012NatPhysMnSiJc}. The $J_c$ value is sample dependent as measurements on MnSi nanowires showed $J_c\sim 22-72$\,MA/m$^2$ \cite{dong2015}, and experiments using ultrasound to detect skyrmion movement found lower values $J_c\sim 50$\,kA/m$^2$ in MnSi single crystals\cite{YK_arxiv}. 
In FeGe, the critical current density has only been measured by direct observation of skyrmion motion using TEM on a 100\,nm thin lamella with $J_c\sim 50$\,kA/m$^2$ at $T=270$\,K.\cite{Yu2012NatComFeGe_Jc}

In Fig.~\ref{figE} we show $\Delta \rho_{yx}(T)$ for different current densities. At 45\,mT, in Fig.~\ref{figE}.a, we show results outside the skyrmion area to test for Joule heating from the contacts or sample. At this field only the local minimum is present and no significant changes occur either in the position or amplitude of the minimum for current densities between 0.65 and 39\,MA/m$^2$. Close inspection of the data reveals a small 10\,mK heating at 39 MA/m$^2$ with respect to 3.9 MA/m$^2$, indicating the absence of any significant heating.
In Fig.~\ref{figE}.b, we show $\Delta \rho_{yx}$ at 276\,K for current densities ranging from 6.5 to 26\,MA/m$^2$. At this temperature, the field sweep encompasses several phases. Starting at low fields, one can see the change of curvature in $\Delta \rho_{yx}(H)$ corresponding to the helical phase below 15\,mT, as well as the  local minimum at 40\,mT, and the THE caused by skyrmions at 75\,mT. We observe no significant changes in neither of these signatures. Data at 3.9\,MA/m$^2$ might indicate a 0.5\,n$\Omega$.cm increase of the THE peak at 75\,mT with respect to higher current densities, but this is barely above noise level. Our experimental resolution of $\approx 0.8$\,nV/$\sqrt{\mathrm{Hz}}$ is insufficient to measure the THE below $\approx 1$\,MA/m$^2$ as in such a small sample (10$\mu$m width) it translates to $0.1\,$nV. 
Despite not seeing important changes, based on the results on FeGe lamellae, the current densities of our experiment most likely place skyrmions in a flow regime. Being in the flow regime implies that the THE of $+5\,$n$\Omega$.cm at 13\,MA/m$^2$ should increase at lower current densities. Therefore, FeGe would have a larger THE than that of MnSi in the skyrmion flow regime ($\approx+2\,$n$\Omega$.cm).  

\section*{Conclusions}
We report the first unambiguous observation of a skyrmion lattice Topological Hall Effect near room temperature in bulk FeGe, using high resolution Hall Effect measurements in a mesoscopic FIB lamella extracted from a single crystal. 
The THE is present in a region of field and temperature close to $T_N$,  where a skyrmion lattice has been observed in bulk samples using SANS\cite{Moskvin2013PRLSANSFeGe}. The amplitude of the THE is of the same magnitude as that of MnSi, and could be even larger at lower current densities. 
We argue that the large THE indicates that the smaller effective magnetic field produced by the bigger FeGe skyrmions is compensated by a larger ordinary Hall constant.
We also report the signature of a reentrant magnetic phase adjacent to the skyrmion phase and connected to the helicoid or inhomogeneous chiral spin state, which confirms that the phase diagram in FeGe is more complex than that of a archetypal skyrmion compound like MnSi.
Finally, FeGe THE magnitude is very similar to that of MnSi, but it is detected in conditions that are more suitable for applications, near room temperature (276\,K) and at lower magnetic fields ($\sim 70$\,mT).

\section*{Methods}

\subsection*{Crystal Growth}
{\small Cubic FeGe is a chiral stoichiometric binary B20 compound, crystallizing in the non-centrosymmetric $P2_13$ cubic space group, isostructural to MnSi. Small single crystals ($\approx100$\,\textmu{}m wide) were grown by chemical vapor transport as described in Ref.~\cite{Richardson1967FeGeCVT}.}

\subsection*{Nanofabrication}
{\small A lamella 0.75\,\textmu{}m thick, 10\,\textmu{}m wide and 25\,\textmu{}m long, was carved out of a pyramidal single crystal, using a FEI Helios 600 FIB/SEM. This lamella is shown in Figure.\ref{figB}.a. We used the known crystallographic orientation of pyramidal single crystals of FeGe, namely the [111] triangular facets with [110] edges, to align the cuts so that the lamella is perpendicular to the [100] axis and its edges are along [110] axes. The lamella was then laid flat on a custom chip, and electrical contacts were made using standard FIB platinum deposition. 2-wire contact resistances ranged from 33\,$\Omega$ for the current contacts, to 40 and 48\,$\Omega$ for voltage contacts, a significant part of which comes from the intrinsically high resistivity of the FIB deposited platinum (1-2\,m$\Omega$.cm per the manufacturer). The result is a lamella for which the magnetic field is applied along [100] and the electrical current is flowing along [110].}

\subsection*{Electrical Transport Measurements}
{\small
We measured the longitudinal resistivity $\rho_{xx}$ and the transverse or Hall resistivity $\rho_{yx}$ in a Quantum Design PPMS with a 9\,T vertical magnet, using a standard 4-wire procedure with a Linear-Research LR-700 AC resistance bridge functioning at a fixed frequency of 15.9\,Hz. We used the same sign convention as in Ref.~\cite{Neubauer2009PRL_THE_MnSi,Huang2012PRL_FeGe_rho_film,Porter2014PRBFeGefilmTransport}, that $\rho_{yx}$ is positive for electrons dominated transport. This combination of sample and setup achieved a noise level of $\approx 0.8$\,nV/$\sqrt{\mathrm{Hz}}$, which is close to the resolution limit of the instrument ($\approx 0.6$\,nV/$\sqrt{\mathrm{Hz}}$) and the intrinsic Johnson-Nyquist noise of the resistive platinum contacts ($\approx 0.7$\,nV/$\sqrt{\mathrm{Hz}}$). Considering measurement duration and current density ($>1\,$MA/m$^2$), this translates into a noise level of typically less than 0.5\,n$\Omega$.cm for the Hall resistivity $\rho_{yx}$, and 15\,n$\Omega$.cm for $\rho_{xx}$, on a sample that is only $0.75\times10\times25$\,\textmu{}m$^3$ in size.

The temperature dependence curves, $\rho_{yx}(T)$, were measured by ramping up the temperature from 270 to 290\,K for several applied magnetic fields, following a zero field cooled (ZFC) procedure. We used the standard magnetic field anti-symmetrization procedure to remove the experimental offset. The offset measured is equivalent to a misalignment of the contacts of $\approx3$\,\textmu{}m. $\rho_{yx}(H)$ curves were measured by cooling in zero field to the target temperature, raising the magnetic field, and then measuring $\rho_{yx}$ while stepping down the field to the opposite value. The field anti-symmetrization procedure was again used to remove the experimental offset. A complete hysteresis cycle was also measured at 276\,K (see SI); compared to the $\rho_{yx}(H)$ curves, the only change in the initial ZFC ramp is an increase (from 15 to 30\,mT) in the upper bound of the helical phase. An extension of the helical phase in magnetic field, has been observed in ZFC magnetization measurements in Fe and Co doped MnSi\cite{Bauer2010PRBMnFeSiMnCoSi}. 
}

\subsection*{Factors in the THE estimate}
The polarization factor is $P = m_s/\mu_{sat}$, where $m_s$ is the ordered magnetic moment (spontaneous magnetization) in the skyrmion phase inferred from a linear extrapolation of the high-field data to H = 0 in an Arrott plot, and $\mu_{sat}$ is the saturated moment deduced from the Curie-Weiss law in the paramagnetic state above $T_N$\cite{Neubauer2009PRLTHE_MnSi}. Ref.~\cite{Porter2014PRBFeGefilmTransport} indicates that $\mu_{sat} = 2.8\pm0.1\, \mu_B/\mathrm{Fe}$ while Ref.~\cite{Wilhelm2016PRB_FeGescaling} finds that $m_s\approx10\,\mathrm{emu}/\mathrm{g} = 0.23\,\mu_B/\mathrm{Fe}$ at 276\,K. We thus estimate the polarization factor in FeGe to be $P \approx 0.23/2.8 = 0.082$, very similar to the value 0.09 in MnSi\cite{Ritz2013PRB_THE_pressure,Neubauer2009PRL_THE_MnSi}.

The effective magnetic field  $B_{eff}$ induced by skyrmions magnetic texture is\cite{Ritz2013PRB_THE_pressure,Neubauer2009PRL_THE_MnSi}:
\begin{equation}
B_{eff} = -\frac{h}{e}\left(\frac{\sqrt{3}}{2\lambda^2_S}\right)
\end{equation}
As the helical period of FeGe is $\lambda_S=69.8$\,nm just below $T_N$\cite{Lebech1989JPCM_FeGe_neutron}, the corresponding effective magnetic field is $B_{eff} = -0.74$\,T, which is 18 times smaller than $B_{eff} = -13.15$\,T in MnSi\cite{Ritz2013PRB_THE_pressure} ($\lambda_S=16.5$\,nm).

Comparing $R_0$ values just above $T_N$, we find $R_0 = +0.11\, \mu\Omega\mathrm{.cm/T}$ in MnSi at 35\,K\cite{Neubauer2009PRL_THE_MnSi} and $R_0 = -0.73\, \mu\Omega\mathrm{.cm/T}$ in our sample of FeGe at 290\,K. Our value of $R_0$ is very close to the value $R_0=-1.09\, \mu\Omega\mathrm{.cm/T}$ in thin films of FeGe\cite{Porter2014PRBFeGefilmTransport}. $R_0$ is thus at least 6 times greater in FeGe than in MnSi.

\subsection*{Data Availability}
{\small
The datasets generated during and/or analysed during the current study are available from the corresponding author on reasonable request.
}
\section*{Acknowledgements}
{\small
B.M. and M.L. are grateful to M.B. Salamon, S.Z. Lin and C. Batista for fruitful discussions.
C.R. thanks C.J.O. Reichhardt for a careful reading.

M.L., B.M and C.R. work was supported by the Laboratory Directed Research and Development program of Los Alamos National Laboratory under project number 20160369ER.
This work was performed, in part, at the Center for Integrated Nanotechnologies, an Office of Science User Facility operated for the U.S. Department of Energy (DOE) Office of Science.
Sandia National Laboratories is a multi-mission laboratory managed and operated by National Technology and Engineering Solutions of Sandia, LLC., a wholly owned subsidiary of Honeywell International, Inc., for the U.S. Department of Energy's National Nuclear Security Administration under contract DE-NA-0003525.
Los Alamos National Laboratory, an affirmative action equal opportunity employer, is operated by Los Alamos National Security, LLC, for the National Nuclear Security Administration of the U.S. Department of Energy under contract DE-AC52-06NA25396.
M.J.S. and S.J. thank the support by NSF grant ECCS-1609585. M.J.S. also acknowledges support from the NSF Graduate Research Fellowship Program grant number DGE-1256259.
}
\section*{Author contributions}
{\small
B.M., S.J. and M.L. planned the experiment. M.J.S and S.J. grew the material. M.L. and D.V.P. prepared the lamella. M.L. performed the measurements. M.L. and B.M. performed the data analysis. M.L., B.M. and C.R. wrote the manuscript and all authors contributed to the final version.
}
\section*{Competing financial interests}
{\small
The authors declare no competing financial interests.
}	
\section*{Materials \& Correspondence}
{\small
Correspondence should be addressed to Dr. Boris Maiorov (maiorov@lanl.gov). Materials request should be addressed to Dr. Song Jin (jin@chem.wisc.edu)
}


\begin{thebibliography}{55}%
\makeatletter
\providecommand \@ifxundefined [1]{%
 \@ifx{#1\undefined}
}%
\providecommand \@ifnum [1]{%
 \ifnum #1\expandafter \@firstoftwo
 \else \expandafter \@secondoftwo
 \fi
}%
\providecommand \@ifx [1]{%
 \ifx #1\expandafter \@firstoftwo
 \else \expandafter \@secondoftwo
 \fi
}%
\providecommand \natexlab [1]{#1}%
\providecommand \enquote  [1]{``#1''}%
\providecommand \bibnamefont  [1]{#1}%
\providecommand \bibfnamefont [1]{#1}%
\providecommand \citenamefont [1]{#1}%
\providecommand \href@noop [0]{\@secondoftwo}%
\providecommand \href [0]{\begingroup \@sanitize@url \@href}%
\providecommand \@href[1]{\@@startlink{#1}\@@href}%
\providecommand \@@href[1]{\endgroup#1\@@endlink}%
\providecommand \@sanitize@url [0]{\catcode `\\12\catcode `\$12\catcode
  `\&12\catcode `\#12\catcode `\^12\catcode `\_12\catcode `\%12\relax}%
\providecommand \@@startlink[1]{}%
\providecommand \@@endlink[0]{}%
\providecommand \url  [0]{\begingroup\@sanitize@url \@url }%
\providecommand \@url [1]{\endgroup\@href {#1}{\urlprefix }}%
\providecommand \urlprefix  [0]{URL }%
\providecommand \Eprint [0]{\href }%
\providecommand \doibase [0]{http://dx.doi.org/}%
\providecommand \selectlanguage [0]{\@gobble}%
\providecommand \bibinfo  [0]{\@secondoftwo}%
\providecommand \bibfield  [0]{\@secondoftwo}%
\providecommand \translation [1]{[#1]}%
\providecommand \BibitemOpen [0]{}%
\providecommand \bibitemStop [0]{}%
\providecommand \bibitemNoStop [0]{.\EOS\space}%
\providecommand \EOS [0]{\spacefactor3000\relax}%
\providecommand \BibitemShut  [1]{\csname bibitem#1\endcsname}%
\let\auto@bib@innerbib\@empty
\bibitem [{\citenamefont {Nagaosa}\ and\ \citenamefont
  {Tokura}(2013)}]{Nagaosa2013NatNanoReview}%
  \BibitemOpen
  \bibfield  {author} {\bibinfo {author} {\bibfnamefont {Naoto}\ \bibnamefont
  {Nagaosa}}\ and\ \bibinfo {author} {\bibfnamefont {Yoshinori}\ \bibnamefont
  {Tokura}},\ }\bibfield  {title} {\enquote {\bibinfo {title} {{Topological
  properties and dynamics of magnetic skyrmions}},}\ }\href {\doibase
  10.1038/nnano.2013.243} {\bibfield  {journal} {\bibinfo  {journal} {Nature
  Nanotechnology}\ }\textbf {\bibinfo {volume} {8}},\ \bibinfo {pages}
  {899--911} (\bibinfo {year} {2013})}\BibitemShut {NoStop}%
\bibitem [{\citenamefont {Fert}\ \emph {et~al.}(2017)\citenamefont {Fert},
  \citenamefont {Reyren},\ and\ \citenamefont
  {Cros}}]{Fert_Reyren_Cros_2017NatRevMat}%
  \BibitemOpen
  \bibfield  {author} {\bibinfo {author} {\bibfnamefont {Albert}\ \bibnamefont
  {Fert}}, \bibinfo {author} {\bibfnamefont {Nicolas}\ \bibnamefont {Reyren}},
  \ and\ \bibinfo {author} {\bibfnamefont {Vincent}\ \bibnamefont {Cros}},\
  }\bibfield  {title} {\enquote {\bibinfo {title} {{Magnetic skyrmions:
  advances in physics and potential applications}},}\ }\href {\doibase
  10.1038/natrevmats.2017.31} {\bibfield  {journal} {\bibinfo  {journal}
  {Nature Reviews Materials}\ }\textbf {\bibinfo {volume} {2}},\ \bibinfo
  {pages} {17031} (\bibinfo {year} {2017})}\BibitemShut {NoStop}%
\bibitem [{\citenamefont {Fert}\ \emph {et~al.}(2013)\citenamefont {Fert},
  \citenamefont {Cros},\ and\ \citenamefont {Sampaio}}]{Fert2013}%
  \BibitemOpen
  \bibfield  {author} {\bibinfo {author} {\bibfnamefont {A}~\bibnamefont
  {Fert}}, \bibinfo {author} {\bibfnamefont {V}~\bibnamefont {Cros}}, \ and\
  \bibinfo {author} {\bibfnamefont {J}~\bibnamefont {Sampaio}},\ }\bibfield
  {title} {\enquote {\bibinfo {title} {{Skyrmions on the track}},}\ }\href
  {\doibase 10.1038/nnano.2013.29} {\bibfield  {journal} {\bibinfo  {journal}
  {Nature Nanotechnology}\ }\textbf {\bibinfo {volume} {8}},\ \bibinfo {pages}
  {152--156} (\bibinfo {year} {2013})}\BibitemShut {NoStop}%
\bibitem [{\citenamefont {Finocchio}\ \emph {et~al.}(2016)\citenamefont
  {Finocchio}, \citenamefont {B{\"{u}}ttner}, \citenamefont {Tomasello},
  \citenamefont {Carpentieri},\ and\ \citenamefont
  {Kl{\"{a}}ui}}]{Finocchio2016JoPDReview}%
  \BibitemOpen
  \bibfield  {author} {\bibinfo {author} {\bibfnamefont {Giovanni}\
  \bibnamefont {Finocchio}}, \bibinfo {author} {\bibfnamefont {Felix}\
  \bibnamefont {B{\"{u}}ttner}}, \bibinfo {author} {\bibfnamefont {Riccardo}\
  \bibnamefont {Tomasello}}, \bibinfo {author} {\bibfnamefont {Mario}\
  \bibnamefont {Carpentieri}}, \ and\ \bibinfo {author} {\bibfnamefont
  {Mathias}\ \bibnamefont {Kl{\"{a}}ui}},\ }\bibfield  {title} {\enquote
  {\bibinfo {title} {{Magnetic skyrmions: from fundamental to applications}},}\
  }\href {\doibase 10.1088/0022-3727/49/42/423001} {\bibfield  {journal}
  {\bibinfo  {journal} {Journal of Physics D: Applied Physics}\ }\textbf
  {\bibinfo {volume} {49}},\ \bibinfo {pages} {423001} (\bibinfo {year}
  {2016})}\BibitemShut {NoStop}%
\bibitem [{\citenamefont {Liu}\ and\ \citenamefont
  {Li}(2015)}]{Liu2015ChinPhysBReview}%
  \BibitemOpen
  \bibfield  {author} {\bibinfo {author} {\bibfnamefont {Ye-Hua}\ \bibnamefont
  {Liu}}\ and\ \bibinfo {author} {\bibfnamefont {You-Quan}\ \bibnamefont
  {Li}},\ }\bibfield  {title} {\enquote {\bibinfo {title} {{Dynamics of
  magnetic skyrmions}},}\ }\href {\doibase 10.1088/1674-1056/24/1/017506}
  {\bibfield  {journal} {\bibinfo  {journal} {Chinese Physics B}\ }\textbf
  {\bibinfo {volume} {24}},\ \bibinfo {pages} {017506} (\bibinfo {year}
  {2015})}\BibitemShut {NoStop}%
\bibitem [{\citenamefont {Belavin}\ and\ \citenamefont
  {Polyakov}(1975)}]{belavin1975skyrmions}%
  \BibitemOpen
  \bibfield  {author} {\bibinfo {author} {\bibfnamefont {AA}~\bibnamefont
  {Belavin}}\ and\ \bibinfo {author} {\bibfnamefont {AM}~\bibnamefont
  {Polyakov}},\ }\bibfield  {title} {\enquote {\bibinfo {title} {Metastable
  states of two-dimensional isotropic ferromagnets},}\ }\href
  {http://www.jetpletters.ac.ru/ps/1529/article_23383.shtml} {\bibfield
  {journal} {\bibinfo  {journal} {Journal of Experimental and Theoretical
  Physics Letters}\ }\textbf {\bibinfo {volume} {22}},\ \bibinfo {pages}
  {245--248} (\bibinfo {year} {1975})}\BibitemShut {NoStop}%
\bibitem [{\citenamefont {Bogdanov}\ and\ \citenamefont
  {Yablonskii}(1989)}]{bogdanov1989skyrmions}%
  \BibitemOpen
  \bibfield  {author} {\bibinfo {author} {\bibfnamefont {AN}~\bibnamefont
  {Bogdanov}}\ and\ \bibinfo {author} {\bibfnamefont {DA}~\bibnamefont
  {Yablonskii}},\ }\bibfield  {title} {\enquote {\bibinfo {title}
  {Thermodynamically stable “vortices” in magnetically ordered crystals.
  the mixed state of magnets},}\ }\href
  {http://www.jetp.ac.ru/cgi-bin/e/index/e/68/1/p101?a=list} {\bibfield
  {journal} {\bibinfo  {journal} {Journal of Experimental and Theoretical
  Physics}\ }\textbf {\bibinfo {volume} {95}},\ \bibinfo {pages} {178}
  (\bibinfo {year} {1989})}\BibitemShut {NoStop}%
\bibitem [{\citenamefont {R{\"{o}}{\ss}ler}\ \emph {et~al.}(2006)\citenamefont
  {R{\"{o}}{\ss}ler}, \citenamefont {Bogdanov},\ and\ \citenamefont
  {Pfleiderer}}]{Roessler2006}%
  \BibitemOpen
  \bibfield  {author} {\bibinfo {author} {\bibfnamefont {U.~K.}\ \bibnamefont
  {R{\"{o}}{\ss}ler}}, \bibinfo {author} {\bibfnamefont {A.~N.}\ \bibnamefont
  {Bogdanov}}, \ and\ \bibinfo {author} {\bibfnamefont {C.}~\bibnamefont
  {Pfleiderer}},\ }\bibfield  {title} {\enquote {\bibinfo {title} {{Spontaneous
  skyrmion ground states in magnetic metals}},}\ }\href {\doibase
  10.1038/nature05056} {\bibfield  {journal} {\bibinfo  {journal} {Nature}\
  }\textbf {\bibinfo {volume} {442}},\ \bibinfo {pages} {797--801} (\bibinfo
  {year} {2006})}\BibitemShut {NoStop}%
\bibitem [{\citenamefont {Muhlbauer}\ \emph {et~al.}(2009)\citenamefont
  {Muhlbauer}, \citenamefont {Binz}, \citenamefont {Jonietz}, \citenamefont
  {Pfleiderer}, \citenamefont {Rosch}, \citenamefont {Neubauer}, \citenamefont
  {Georgii},\ and\ \citenamefont {Boni}}]{Muhlbauer2009ScienceMnSineutron}%
  \BibitemOpen
  \bibfield  {author} {\bibinfo {author} {\bibfnamefont {S.}~\bibnamefont
  {Muhlbauer}}, \bibinfo {author} {\bibfnamefont {B.}~\bibnamefont {Binz}},
  \bibinfo {author} {\bibfnamefont {F.}~\bibnamefont {Jonietz}}, \bibinfo
  {author} {\bibfnamefont {C}~\bibnamefont {Pfleiderer}}, \bibinfo {author}
  {\bibfnamefont {A}~\bibnamefont {Rosch}}, \bibinfo {author} {\bibfnamefont
  {A}~\bibnamefont {Neubauer}}, \bibinfo {author} {\bibfnamefont {Robert}\
  \bibnamefont {Georgii}}, \ and\ \bibinfo {author} {\bibfnamefont
  {P.}~\bibnamefont {Boni}},\ }\bibfield  {title} {\enquote {\bibinfo {title}
  {{Skyrmion Lattice in a Chiral Magnet}},}\ }\href {\doibase
  10.1126/science.1166767} {\bibfield  {journal} {\bibinfo  {journal}
  {Science}\ }\textbf {\bibinfo {volume} {323}},\ \bibinfo {pages} {915--919}
  (\bibinfo {year} {2009})}\BibitemShut {NoStop}%
\bibitem [{\citenamefont {Yu}\ \emph {et~al.}(2011)\citenamefont {Yu},
  \citenamefont {Kanazawa}, \citenamefont {Onose}, \citenamefont {Kimoto},
  \citenamefont {Zhang}, \citenamefont {Ishiwata}, \citenamefont {Matsui},\
  and\ \citenamefont {Tokura}}]{Yu2011NatMat}%
  \BibitemOpen
  \bibfield  {author} {\bibinfo {author} {\bibfnamefont {X~Z}\ \bibnamefont
  {Yu}}, \bibinfo {author} {\bibfnamefont {N}~\bibnamefont {Kanazawa}},
  \bibinfo {author} {\bibfnamefont {Y}~\bibnamefont {Onose}}, \bibinfo {author}
  {\bibfnamefont {K}~\bibnamefont {Kimoto}}, \bibinfo {author} {\bibfnamefont
  {W~Z}\ \bibnamefont {Zhang}}, \bibinfo {author} {\bibfnamefont
  {S}~\bibnamefont {Ishiwata}}, \bibinfo {author} {\bibfnamefont
  {Y}~\bibnamefont {Matsui}}, \ and\ \bibinfo {author} {\bibfnamefont
  {Y}~\bibnamefont {Tokura}},\ }\bibfield  {title} {\enquote {\bibinfo {title}
  {{Near room-temperature formation of a skyrmion crystal in thin-films of the
  helimagnet FeGe.}}}\ }\href {\doibase 10.1038/nmat2916} {\bibfield  {journal}
  {\bibinfo  {journal} {Nature materials}\ }\textbf {\bibinfo {volume} {10}},\
  \bibinfo {pages} {106--109} (\bibinfo {year} {2011})}\BibitemShut {NoStop}%
\bibitem [{\citenamefont {Boulle}\ \emph {et~al.}(2016)\citenamefont {Boulle},
  \citenamefont {Vogel}, \citenamefont {Yang}, \citenamefont {Pizzini},
  \citenamefont {{de Souza Chaves}}, \citenamefont {Locatelli}, \citenamefont
  {Menteş}, \citenamefont {Sala}, \citenamefont {Buda-Prejbeanu},
  \citenamefont {Klein}, \citenamefont {Belmeguenai}, \citenamefont
  {Roussign{\'{e}}}, \citenamefont {Stashkevich}, \citenamefont {Ch{\'{e}}rif},
  \citenamefont {Aballe}, \citenamefont {Foerster}, \citenamefont {Chshiev},
  \citenamefont {Auffret}, \citenamefont {Miron},\ and\ \citenamefont
  {Gaudin}}]{Boulle2016NatNano}%
  \BibitemOpen
  \bibfield  {author} {\bibinfo {author} {\bibfnamefont {Olivier}\ \bibnamefont
  {Boulle}}, \bibinfo {author} {\bibfnamefont {Jan}\ \bibnamefont {Vogel}},
  \bibinfo {author} {\bibfnamefont {Hongxin}\ \bibnamefont {Yang}}, \bibinfo
  {author} {\bibfnamefont {Stefania}\ \bibnamefont {Pizzini}}, \bibinfo
  {author} {\bibfnamefont {Dayane}\ \bibnamefont {{de Souza Chaves}}}, \bibinfo
  {author} {\bibfnamefont {Andrea}\ \bibnamefont {Locatelli}}, \bibinfo
  {author} {\bibfnamefont {Tevfik~Onur}\ \bibnamefont {Menteş}}, \bibinfo
  {author} {\bibfnamefont {Alessandro}\ \bibnamefont {Sala}}, \bibinfo {author}
  {\bibfnamefont {Liliana~D.}\ \bibnamefont {Buda-Prejbeanu}}, \bibinfo
  {author} {\bibfnamefont {Olivier}\ \bibnamefont {Klein}}, \bibinfo {author}
  {\bibfnamefont {Mohamed}\ \bibnamefont {Belmeguenai}}, \bibinfo {author}
  {\bibfnamefont {Yves}\ \bibnamefont {Roussign{\'{e}}}}, \bibinfo {author}
  {\bibfnamefont {Andrey}\ \bibnamefont {Stashkevich}}, \bibinfo {author}
  {\bibfnamefont {Salim~Mourad}\ \bibnamefont {Ch{\'{e}}rif}}, \bibinfo
  {author} {\bibfnamefont {Lucia}\ \bibnamefont {Aballe}}, \bibinfo {author}
  {\bibfnamefont {Michael}\ \bibnamefont {Foerster}}, \bibinfo {author}
  {\bibfnamefont {Mairbek}\ \bibnamefont {Chshiev}}, \bibinfo {author}
  {\bibfnamefont {St{\'{e}}phane}\ \bibnamefont {Auffret}}, \bibinfo {author}
  {\bibfnamefont {Ioan~Mihai}\ \bibnamefont {Miron}}, \ and\ \bibinfo {author}
  {\bibfnamefont {Gilles}\ \bibnamefont {Gaudin}},\ }\href {\doibase
  10.1038/nnano.2015.315} {\bibfield  {journal} {\bibinfo  {journal} {Nature
  Nanotechnology}\ }\textbf {\bibinfo {volume} {11}},\ \bibinfo {pages}
  {449--454} (\bibinfo {year} {2016})},\ \Eprint
  {http://arxiv.org/abs/1601.02278} {1601.02278} \BibitemShut {NoStop}%
\bibitem [{\citenamefont {Dzialoshinskii}(1957)}]{dzialoshinskii1957}%
  \BibitemOpen
  \bibfield  {author} {\bibinfo {author} {\bibfnamefont {IE}~\bibnamefont
  {Dzialoshinskii}},\ }\bibfield  {title} {\enquote {\bibinfo {title}
  {Thermodynamic theory of weak ferromagnetism in antiferromagnetic
  substances},}\ }\href
  {http://www.jetp.ac.ru/cgi-bin/e/index/e/5/6/p1259?a=list} {\bibfield
  {journal} {\bibinfo  {journal} {Journal of Experimental and Theoretical
  Physics}\ }\textbf {\bibinfo {volume} {5}},\ \bibinfo {pages} {1259--1272}
  (\bibinfo {year} {1957})}\BibitemShut {NoStop}%
\bibitem [{\citenamefont {Moriya}(1960)}]{Moriya1960}%
  \BibitemOpen
  \bibfield  {author} {\bibinfo {author} {\bibfnamefont {T\^oru}\ \bibnamefont
  {Moriya}},\ }\bibfield  {title} {\enquote {\bibinfo {title} {Anisotropic
  superexchange interaction and weak ferromagnetism},}\ }\href {\doibase
  10.1103/PhysRev.120.91} {\bibfield  {journal} {\bibinfo  {journal} {Phys.
  Rev.}\ }\textbf {\bibinfo {volume} {120}},\ \bibinfo {pages} {91--98}
  (\bibinfo {year} {1960})}\BibitemShut {NoStop}%
\bibitem [{\citenamefont {Braun}(2012)}]{Braun2012}%
  \BibitemOpen
  \bibfield  {author} {\bibinfo {author} {\bibfnamefont {Hans-Benjamin}\
  \bibnamefont {Braun}},\ }\bibfield  {title} {\enquote {\bibinfo {title}
  {{Topological effects in nanomagnetism: from superparamagnetism to chiral
  quantum solitons}},}\ }\href {\doibase 10.1080/00018732.2012.663070}
  {\bibfield  {journal} {\bibinfo  {journal} {Advances in Physics}\ }\textbf
  {\bibinfo {volume} {61}},\ \bibinfo {pages} {1--116} (\bibinfo {year}
  {2012})}\BibitemShut {NoStop}%
\bibitem [{\citenamefont {Jonietz}\ \emph {et~al.}(2011)\citenamefont
  {Jonietz}, \citenamefont {Mulbauer}, \citenamefont {Pfleiderer},
  \citenamefont {Neubauer}, \citenamefont {Munzer}, \citenamefont {Bauer},
  \citenamefont {Adams}, \citenamefont {Georgii}, \citenamefont {B{\"{o}}ni},
  \citenamefont {Duine}, \citenamefont {Everschor}, \citenamefont {Garst},\
  and\ \citenamefont {Rosch}}]{Jonietz2011Science}%
  \BibitemOpen
  \bibfield  {author} {\bibinfo {author} {\bibfnamefont {F.}~\bibnamefont
  {Jonietz}}, \bibinfo {author} {\bibfnamefont {S.}~\bibnamefont {Mulbauer}},
  \bibinfo {author} {\bibfnamefont {C.}~\bibnamefont {Pfleiderer}}, \bibinfo
  {author} {\bibfnamefont {A.}~\bibnamefont {Neubauer}}, \bibinfo {author}
  {\bibfnamefont {W.}~\bibnamefont {Munzer}}, \bibinfo {author} {\bibfnamefont
  {A.}~\bibnamefont {Bauer}}, \bibinfo {author} {\bibfnamefont
  {T.}~\bibnamefont {Adams}}, \bibinfo {author} {\bibfnamefont
  {R.}~\bibnamefont {Georgii}}, \bibinfo {author} {\bibfnamefont
  {P.}~\bibnamefont {B{\"{o}}ni}}, \bibinfo {author} {\bibfnamefont {R.~A.}\
  \bibnamefont {Duine}}, \bibinfo {author} {\bibfnamefont {K.}~\bibnamefont
  {Everschor}}, \bibinfo {author} {\bibfnamefont {M.}~\bibnamefont {Garst}}, \
  and\ \bibinfo {author} {\bibfnamefont {A.}~\bibnamefont {Rosch}},\ }\bibfield
   {title} {\enquote {\bibinfo {title} {{Spin Transfer Torques in MnSi}},}\
  }\href {\doibase 10.1126/science.1195709} {\bibfield  {journal} {\bibinfo
  {journal} {Science}\ ,\ \bibinfo {pages} {1648--1652}} (\bibinfo {year}
  {2011})}\BibitemShut {NoStop}%
\bibitem [{\citenamefont {Schulz}\ \emph {et~al.}(2012)\citenamefont {Schulz},
  \citenamefont {Ritz}, \citenamefont {Bauer}, \citenamefont {Halder},
  \citenamefont {Wagner}, \citenamefont {Franz}, \citenamefont {Pfleiderer},
  \citenamefont {Everschor}, \citenamefont {Garst},\ and\ \citenamefont
  {Rosch}}]{Schulz2012NatPhysMnSiJc}%
  \BibitemOpen
  \bibfield  {author} {\bibinfo {author} {\bibfnamefont {T.}~\bibnamefont
  {Schulz}}, \bibinfo {author} {\bibfnamefont {R.}~\bibnamefont {Ritz}},
  \bibinfo {author} {\bibfnamefont {A.}~\bibnamefont {Bauer}}, \bibinfo
  {author} {\bibfnamefont {M.}~\bibnamefont {Halder}}, \bibinfo {author}
  {\bibfnamefont {M.}~\bibnamefont {Wagner}}, \bibinfo {author} {\bibfnamefont
  {C.}~\bibnamefont {Franz}}, \bibinfo {author} {\bibfnamefont
  {C.}~\bibnamefont {Pfleiderer}}, \bibinfo {author} {\bibfnamefont
  {K.}~\bibnamefont {Everschor}}, \bibinfo {author} {\bibfnamefont
  {M.}~\bibnamefont {Garst}}, \ and\ \bibinfo {author} {\bibfnamefont
  {A.}~\bibnamefont {Rosch}},\ }\bibfield  {title} {\enquote {\bibinfo {title}
  {{Emergent electrodynamics of skyrmions in a chiral magnet}},}\ }\href
  {\doibase 10.1038/nphys2231} {\bibfield  {journal} {\bibinfo  {journal}
  {Nature Physics}\ }\textbf {\bibinfo {volume} {8}},\ \bibinfo {pages}
  {301--304} (\bibinfo {year} {2012})}\BibitemShut {NoStop}%
\bibitem [{\citenamefont {Reichhardt}\ \emph {et~al.}(2015)\citenamefont
  {Reichhardt}, \citenamefont {Ray},\ and\ \citenamefont
  {Reichhardt}}]{reichhardt1}%
  \BibitemOpen
  \bibfield  {author} {\bibinfo {author} {\bibfnamefont {C.}~\bibnamefont
  {Reichhardt}}, \bibinfo {author} {\bibfnamefont {D.}~\bibnamefont {Ray}}, \
  and\ \bibinfo {author} {\bibfnamefont {C.~J.~Olson}\ \bibnamefont
  {Reichhardt}},\ }\bibfield  {title} {\enquote {\bibinfo {title} {{Collective
  Transport Properties of Driven Skyrmions with Random Disorder}},}\ }\href
  {\doibase 10.1103/PhysRevLett.114.217202} {\bibfield  {journal} {\bibinfo
  {journal} {Phys. Rev. Lett.}\ }\textbf {\bibinfo {volume} {114}},\ \bibinfo
  {pages} {217202} (\bibinfo {year} {2015})}\BibitemShut {NoStop}%
\bibitem [{\citenamefont {Iwasaki}\ \emph {et~al.}(2013)\citenamefont
  {Iwasaki}, \citenamefont {Mochizuki},\ and\ \citenamefont
  {Nagaosa}}]{iwasaki2013current}%
  \BibitemOpen
  \bibfield  {author} {\bibinfo {author} {\bibfnamefont {Junichi}\ \bibnamefont
  {Iwasaki}}, \bibinfo {author} {\bibfnamefont {Masahito}\ \bibnamefont
  {Mochizuki}}, \ and\ \bibinfo {author} {\bibfnamefont {Naoto}\ \bibnamefont
  {Nagaosa}},\ }\bibfield  {title} {\enquote {\bibinfo {title} {Current-induced
  skyrmion dynamics in constricted geometries},}\ }\href {\doibase
  10.1038/nnano.2013.176} {\bibfield  {journal} {\bibinfo  {journal} {Nature
  nanotechnology}\ }\textbf {\bibinfo {volume} {8}},\ \bibinfo {pages}
  {742--747} (\bibinfo {year} {2013})}\BibitemShut {NoStop}%
\bibitem [{\citenamefont {Zhang}\ \emph {et~al.}(2015)\citenamefont {Zhang},
  \citenamefont {Ezawa},\ and\ \citenamefont {Zhou}}]{zhang2015magnetic}%
  \BibitemOpen
  \bibfield  {author} {\bibinfo {author} {\bibfnamefont {Xichao}\ \bibnamefont
  {Zhang}}, \bibinfo {author} {\bibfnamefont {Motohiko}\ \bibnamefont {Ezawa}},
  \ and\ \bibinfo {author} {\bibfnamefont {Yan}\ \bibnamefont {Zhou}},\
  }\bibfield  {title} {\enquote {\bibinfo {title} {Magnetic skyrmion logic
  gates: conversion, duplication and merging of skyrmions},}\ }\href {\doibase
  10.1038/srep09400} {\bibfield  {journal} {\bibinfo  {journal} {Scientific
  reports}\ }\textbf {\bibinfo {volume} {5}} (\bibinfo {year} {2015}),\
  10.1038/srep09400}\BibitemShut {NoStop}%
\bibitem [{\citenamefont {Jiang}\ \emph {et~al.}(2015)\citenamefont {Jiang},
  \citenamefont {Upadhyaya}, \citenamefont {Zhang}, \citenamefont {Yu},
  \citenamefont {Jungfleisch}, \citenamefont {Fradin}, \citenamefont {Pearson},
  \citenamefont {Tserkovnyak}, \citenamefont {Wang}, \citenamefont {Heinonen},
  \citenamefont {te~Velthuis},\ and\ \citenamefont
  {Hoffmann}}]{jiang2015blowing}%
  \BibitemOpen
  \bibfield  {author} {\bibinfo {author} {\bibfnamefont {Wanjun}\ \bibnamefont
  {Jiang}}, \bibinfo {author} {\bibfnamefont {Pramey}\ \bibnamefont
  {Upadhyaya}}, \bibinfo {author} {\bibfnamefont {Wei}\ \bibnamefont {Zhang}},
  \bibinfo {author} {\bibfnamefont {Guoqiang}\ \bibnamefont {Yu}}, \bibinfo
  {author} {\bibfnamefont {M.~Benjamin}\ \bibnamefont {Jungfleisch}}, \bibinfo
  {author} {\bibfnamefont {Frank~Y.}\ \bibnamefont {Fradin}}, \bibinfo {author}
  {\bibfnamefont {John~E.}\ \bibnamefont {Pearson}}, \bibinfo {author}
  {\bibfnamefont {Yaroslav}\ \bibnamefont {Tserkovnyak}}, \bibinfo {author}
  {\bibfnamefont {Kang~L.}\ \bibnamefont {Wang}}, \bibinfo {author}
  {\bibfnamefont {Olle}\ \bibnamefont {Heinonen}}, \bibinfo {author}
  {\bibfnamefont {S.~G.~E.}\ \bibnamefont {te~Velthuis}}, \ and\ \bibinfo
  {author} {\bibfnamefont {Axel}\ \bibnamefont {Hoffmann}},\ }\bibfield
  {title} {\enquote {\bibinfo {title} {{Blowing magnetic skyrmion bubbles}},}\
  }\href {\doibase 10.1126/science.aaa1442} {\bibfield  {journal} {\bibinfo
  {journal} {Science}\ }\textbf {\bibinfo {volume} {349}},\ \bibinfo {pages}
  {283--286} (\bibinfo {year} {2015})}\BibitemShut {NoStop}%
\bibitem [{\citenamefont {B{\"u}ttner}\ \emph {et~al.}(2015)\citenamefont
  {B{\"u}ttner}, \citenamefont {Moutafis}, \citenamefont {Schneider},
  \citenamefont {Kr{\"u}ger}, \citenamefont {G{\"u}nther}, \citenamefont
  {Geilhufe}, \citenamefont {Schmising}, \citenamefont {Mohanty}, \citenamefont
  {Pfau}, \citenamefont {Schaffert} \emph {et~al.}}]{buttner2015dynamics}%
  \BibitemOpen
  \bibfield  {author} {\bibinfo {author} {\bibfnamefont {Felix}\ \bibnamefont
  {B{\"u}ttner}}, \bibinfo {author} {\bibfnamefont {C}~\bibnamefont
  {Moutafis}}, \bibinfo {author} {\bibfnamefont {M}~\bibnamefont {Schneider}},
  \bibinfo {author} {\bibfnamefont {B}~\bibnamefont {Kr{\"u}ger}}, \bibinfo
  {author} {\bibfnamefont {CM}~\bibnamefont {G{\"u}nther}}, \bibinfo {author}
  {\bibfnamefont {J}~\bibnamefont {Geilhufe}}, \bibinfo {author} {\bibfnamefont
  {C~v~Korff}\ \bibnamefont {Schmising}}, \bibinfo {author} {\bibfnamefont
  {J}~\bibnamefont {Mohanty}}, \bibinfo {author} {\bibfnamefont
  {B}~\bibnamefont {Pfau}}, \bibinfo {author} {\bibfnamefont {S}~\bibnamefont
  {Schaffert}},  \emph {et~al.},\ }\bibfield  {title} {\enquote {\bibinfo
  {title} {Dynamics and inertia of skyrmionic spin structures},}\ }\href
  {\doibase 10.1038/nphys3234} {\bibfield  {journal} {\bibinfo  {journal}
  {Nature Physics}\ }\textbf {\bibinfo {volume} {11}},\ \bibinfo {pages}
  {225--228} (\bibinfo {year} {2015})}\BibitemShut {NoStop}%
\bibitem [{\citenamefont {Woo}\ \emph {et~al.}(2016)\citenamefont {Woo},
  \citenamefont {Litzius}, \citenamefont {Kr{\"{u}}ger}, \citenamefont {Im},
  \citenamefont {Caretta}, \citenamefont {Richter}, \citenamefont {Mann},
  \citenamefont {Krone}, \citenamefont {Reeve}, \citenamefont {Weigand},
  \citenamefont {Agrawal}, \citenamefont {Lemesh}, \citenamefont {Mawass},
  \citenamefont {Fischer}, \citenamefont {Kl{\"{a}}ui},\ and\ \citenamefont
  {Beach}}]{Woo2016}%
  \BibitemOpen
  \bibfield  {author} {\bibinfo {author} {\bibfnamefont {Seonghoon}\
  \bibnamefont {Woo}}, \bibinfo {author} {\bibfnamefont {Kai}\ \bibnamefont
  {Litzius}}, \bibinfo {author} {\bibfnamefont {Benjamin}\ \bibnamefont
  {Kr{\"{u}}ger}}, \bibinfo {author} {\bibfnamefont {Mi-Young M-Y.}\
  \bibnamefont {Im}}, \bibinfo {author} {\bibfnamefont {Lucas}\ \bibnamefont
  {Caretta}}, \bibinfo {author} {\bibfnamefont {Kornel}\ \bibnamefont
  {Richter}}, \bibinfo {author} {\bibfnamefont {Maxwell}\ \bibnamefont {Mann}},
  \bibinfo {author} {\bibfnamefont {Andrea}\ \bibnamefont {Krone}}, \bibinfo
  {author} {\bibfnamefont {Robert~M}\ \bibnamefont {Reeve}}, \bibinfo {author}
  {\bibfnamefont {Markus}\ \bibnamefont {Weigand}}, \bibinfo {author}
  {\bibfnamefont {Parnika}\ \bibnamefont {Agrawal}}, \bibinfo {author}
  {\bibfnamefont {Ivan}\ \bibnamefont {Lemesh}}, \bibinfo {author}
  {\bibfnamefont {M-A. Mohamad-Assaad}\ \bibnamefont {Mawass}}, \bibinfo
  {author} {\bibfnamefont {Peter}\ \bibnamefont {Fischer}}, \bibinfo {author}
  {\bibfnamefont {Mathias}\ \bibnamefont {Kl{\"{a}}ui}}, \ and\ \bibinfo
  {author} {\bibfnamefont {Geoffrey S~D}\ \bibnamefont {Beach}},\ }\bibfield
  {title} {\enquote {\bibinfo {title} {{Observation of room-temperature
  magnetic skyrmions and their current-driven dynamics in ultrathin metallic
  ferromagnets}},}\ }\href {\doibase 10.1038/nmat4593} {\bibfield  {journal}
  {\bibinfo  {journal} {Nature Materials}\ }\textbf {\bibinfo {volume} {15}},\
  \bibinfo {pages} {501--506} (\bibinfo {year} {2016})}\BibitemShut {NoStop}%
\bibitem [{\citenamefont {Juge}\ \emph {et~al.}(2017)\citenamefont {Juge},
  \citenamefont {Je}, \citenamefont {{de Souza Chaves}}, \citenamefont
  {Pizzini}, \citenamefont {Buda-Prejbeanu}, \citenamefont {Aballe},
  \citenamefont {Foerster}, \citenamefont {Locatelli}, \citenamefont {Menteş},
  \citenamefont {Sala}, \citenamefont {Maccherozzi}, \citenamefont {Dhesi},
  \citenamefont {Auffret}, \citenamefont {Gautier}, \citenamefont {Gaudin},
  \citenamefont {Vogel},\ and\ \citenamefont {Boulle}}]{Juge2017a}%
  \BibitemOpen
  \bibfield  {author} {\bibinfo {author} {\bibfnamefont {Rom{\'{e}}o}\
  \bibnamefont {Juge}}, \bibinfo {author} {\bibfnamefont {Soong-Geun}\
  \bibnamefont {Je}}, \bibinfo {author} {\bibfnamefont {Dayane}\ \bibnamefont
  {{de Souza Chaves}}}, \bibinfo {author} {\bibfnamefont {Stefania}\
  \bibnamefont {Pizzini}}, \bibinfo {author} {\bibfnamefont {Liliana~D.}\
  \bibnamefont {Buda-Prejbeanu}}, \bibinfo {author} {\bibfnamefont {Lucia}\
  \bibnamefont {Aballe}}, \bibinfo {author} {\bibfnamefont {Michael}\
  \bibnamefont {Foerster}}, \bibinfo {author} {\bibfnamefont {Andrea}\
  \bibnamefont {Locatelli}}, \bibinfo {author} {\bibfnamefont {Tevfik~Onur}\
  \bibnamefont {Menteş}}, \bibinfo {author} {\bibfnamefont {Alessandro}\
  \bibnamefont {Sala}}, \bibinfo {author} {\bibfnamefont {Francesco}\
  \bibnamefont {Maccherozzi}}, \bibinfo {author} {\bibfnamefont {Sarnjeet~S.}\
  \bibnamefont {Dhesi}}, \bibinfo {author} {\bibfnamefont {St{\'{e}}phane}\
  \bibnamefont {Auffret}}, \bibinfo {author} {\bibfnamefont {Eric}\
  \bibnamefont {Gautier}}, \bibinfo {author} {\bibfnamefont {Gilles}\
  \bibnamefont {Gaudin}}, \bibinfo {author} {\bibfnamefont {Jan}\ \bibnamefont
  {Vogel}}, \ and\ \bibinfo {author} {\bibfnamefont {Olivier}\ \bibnamefont
  {Boulle}},\ }\bibfield  {title} {\enquote {\bibinfo {title} {{Magnetic
  skyrmions in confined geometries: Effect of the magnetic field and the
  disorder}},}\ }\href {\doibase 10.1016/j.jmmm.2017.10.030} {\bibfield
  {journal} {\bibinfo  {journal} {Journal of Magnetism and Magnetic Materials}\
  } (\bibinfo {year} {2017}),\ 10.1016/j.jmmm.2017.10.030}\BibitemShut
  {NoStop}%
\bibitem [{\citenamefont {Maccariello}\ \emph {et~al.}(2018)\citenamefont
  {Maccariello}, \citenamefont {Legrand}, \citenamefont {Reyren}, \citenamefont
  {Garcia}, \citenamefont {Bouzehouane}, \citenamefont {Collin}, \citenamefont
  {Cros},\ and\ \citenamefont {Fert}}]{Maccariello2018a}%
  \BibitemOpen
  \bibfield  {author} {\bibinfo {author} {\bibfnamefont {Davide}\ \bibnamefont
  {Maccariello}}, \bibinfo {author} {\bibfnamefont {William}\ \bibnamefont
  {Legrand}}, \bibinfo {author} {\bibfnamefont {Nicolas}\ \bibnamefont
  {Reyren}}, \bibinfo {author} {\bibfnamefont {Karin}\ \bibnamefont {Garcia}},
  \bibinfo {author} {\bibfnamefont {Karim}\ \bibnamefont {Bouzehouane}},
  \bibinfo {author} {\bibfnamefont {Sophie}\ \bibnamefont {Collin}}, \bibinfo
  {author} {\bibfnamefont {Vincent}\ \bibnamefont {Cros}}, \ and\ \bibinfo
  {author} {\bibfnamefont {Albert}\ \bibnamefont {Fert}},\ }\bibfield  {title}
  {\enquote {\bibinfo {title} {{Electrical detection of single magnetic
  skyrmions in metallic multilayers at room temperature}},}\ }\href {\doibase
  10.1038/s41565-017-0044-4} {\bibfield  {journal} {\bibinfo  {journal} {Nature
  Nanotechnology}\ }\textbf {\bibinfo {volume} {7}},\ \bibinfo {pages} {056022}
  (\bibinfo {year} {2018})}\BibitemShut {NoStop}%
\bibitem [{\citenamefont {Neubauer}\ \emph
  {et~al.}(2009{\natexlab{a}})\citenamefont {Neubauer}, \citenamefont
  {Pfleiderer}, \citenamefont {Binz}, \citenamefont {Rosch}, \citenamefont
  {Ritz}, \citenamefont {Niklowitz},\ and\ \citenamefont
  {B\"oni}}]{Neubauer2009PRLTHE_MnSi}%
  \BibitemOpen
  \bibfield  {author} {\bibinfo {author} {\bibfnamefont {A.}~\bibnamefont
  {Neubauer}}, \bibinfo {author} {\bibfnamefont {C.}~\bibnamefont
  {Pfleiderer}}, \bibinfo {author} {\bibfnamefont {B.}~\bibnamefont {Binz}},
  \bibinfo {author} {\bibfnamefont {A.}~\bibnamefont {Rosch}}, \bibinfo
  {author} {\bibfnamefont {R.}~\bibnamefont {Ritz}}, \bibinfo {author}
  {\bibfnamefont {P.~G.}\ \bibnamefont {Niklowitz}}, \ and\ \bibinfo {author}
  {\bibfnamefont {P.}~\bibnamefont {B\"oni}},\ }\bibfield  {title} {\enquote
  {\bibinfo {title} {{Topological Hall Effect in the $A$ Phase of MnSi}},}\
  }\href {\doibase 10.1103/PhysRevLett.102.186602} {\bibfield  {journal}
  {\bibinfo  {journal} {Phys. Rev. Lett.}\ }\textbf {\bibinfo {volume} {102}},\
  \bibinfo {pages} {186602} (\bibinfo {year} {2009}{\natexlab{a}})}\BibitemShut
  {NoStop}%
\bibitem [{\citenamefont {Franz}\ \emph {et~al.}(2014)\citenamefont {Franz},
  \citenamefont {Freimuth}, \citenamefont {Bauer}, \citenamefont {Ritz},
  \citenamefont {Schnarr}, \citenamefont {Duvinage}, \citenamefont {Adams},
  \citenamefont {Bl\"ugel}, \citenamefont {Rosch}, \citenamefont {Mokrousov},\
  and\ \citenamefont {Pfleiderer}}]{Franz2014PRL_HallMnCoFeSi}%
  \BibitemOpen
  \bibfield  {author} {\bibinfo {author} {\bibfnamefont {C.}~\bibnamefont
  {Franz}}, \bibinfo {author} {\bibfnamefont {F.}~\bibnamefont {Freimuth}},
  \bibinfo {author} {\bibfnamefont {A.}~\bibnamefont {Bauer}}, \bibinfo
  {author} {\bibfnamefont {R.}~\bibnamefont {Ritz}}, \bibinfo {author}
  {\bibfnamefont {C.}~\bibnamefont {Schnarr}}, \bibinfo {author} {\bibfnamefont
  {C.}~\bibnamefont {Duvinage}}, \bibinfo {author} {\bibfnamefont
  {T.}~\bibnamefont {Adams}}, \bibinfo {author} {\bibfnamefont
  {S.}~\bibnamefont {Bl\"ugel}}, \bibinfo {author} {\bibfnamefont
  {A.}~\bibnamefont {Rosch}}, \bibinfo {author} {\bibfnamefont
  {Y.}~\bibnamefont {Mokrousov}}, \ and\ \bibinfo {author} {\bibfnamefont
  {C.}~\bibnamefont {Pfleiderer}},\ }\bibfield  {title} {\enquote {\bibinfo
  {title} {{Real-Space and Reciprocal-Space Berry Phases in the Hall Effect of
  ${\mathrm{Mn}}_{1\ensuremath{-}x}{\mathrm{Fe}}_{x}\mathrm{Si}$}},}\ }\href
  {\doibase 10.1103/PhysRevLett.112.186601} {\bibfield  {journal} {\bibinfo
  {journal} {Phys. Rev. Lett.}\ }\textbf {\bibinfo {volume} {112}},\ \bibinfo
  {pages} {186601} (\bibinfo {year} {2014})}\BibitemShut {NoStop}%
\bibitem [{\citenamefont {Liang}\ \emph {et~al.}(2015)\citenamefont {Liang},
  \citenamefont {DeGrave}, \citenamefont {Stolt}, \citenamefont {Tokura},\ and\
  \citenamefont {Jin}}]{dong2015}%
  \BibitemOpen
  \bibfield  {author} {\bibinfo {author} {\bibfnamefont {Dong}\ \bibnamefont
  {Liang}}, \bibinfo {author} {\bibfnamefont {John~P.}\ \bibnamefont
  {DeGrave}}, \bibinfo {author} {\bibfnamefont {Matthew~J.}\ \bibnamefont
  {Stolt}}, \bibinfo {author} {\bibfnamefont {Yoshinori}\ \bibnamefont
  {Tokura}}, \ and\ \bibinfo {author} {\bibfnamefont {Song}\ \bibnamefont
  {Jin}},\ }\bibfield  {title} {\enquote {\bibinfo {title} {{Current-driven
  dynamics of skyrmions stabilized in MnSi nanowires revealed by topological
  Hall effect}},}\ }\href {\doibase 10.1038/ncomms9217} {\bibfield  {journal}
  {\bibinfo  {journal} {Nature Communications}\ }\textbf {\bibinfo {volume}
  {6}},\ \bibinfo {pages} {8217} (\bibinfo {year} {2015})}\BibitemShut
  {NoStop}%
\bibitem [{\citenamefont {Tokunaga}\ \emph {et~al.}(2015)\citenamefont
  {Tokunaga}, \citenamefont {Yu}, \citenamefont {White}, \citenamefont
  {R{\o}nnow}, \citenamefont {Morikawa}, \citenamefont {Taguchi},\ and\
  \citenamefont {Tokura}}]{Tokunaga2015NatComCoZnMn}%
  \BibitemOpen
  \bibfield  {author} {\bibinfo {author} {\bibfnamefont {Y}~\bibnamefont
  {Tokunaga}}, \bibinfo {author} {\bibfnamefont {X~Z}\ \bibnamefont {Yu}},
  \bibinfo {author} {\bibfnamefont {J~S}\ \bibnamefont {White}}, \bibinfo
  {author} {\bibfnamefont {H.~M.}\ \bibnamefont {R{\o}nnow}}, \bibinfo {author}
  {\bibfnamefont {D}~\bibnamefont {Morikawa}}, \bibinfo {author} {\bibfnamefont
  {Y}~\bibnamefont {Taguchi}}, \ and\ \bibinfo {author} {\bibfnamefont
  {Y}~\bibnamefont {Tokura}},\ }\bibfield  {title} {\enquote {\bibinfo {title}
  {{A new class of chiral materials hosting magnetic skyrmions beyond room
  temperature}},}\ }\href {\doibase 10.1038/ncomms8638} {\bibfield  {journal}
  {\bibinfo  {journal} {Nature Communications}\ }\textbf {\bibinfo {volume}
  {6}},\ \bibinfo {pages} {7638} (\bibinfo {year} {2015})}\BibitemShut
  {NoStop}%
\bibitem [{\citenamefont {Yu}\ \emph {et~al.}(2017)\citenamefont {Yu},
  \citenamefont {Morikawa}, \citenamefont {Tokunaga}, \citenamefont {Kubota},
  \citenamefont {Kurumaji}, \citenamefont {Oike}, \citenamefont {Nakamura},
  \citenamefont {Kagawa}, \citenamefont {Taguchi}, \citenamefont {hisa Arima},
  \citenamefont {Kawasaki}, \citenamefont {Tokura}, \citenamefont {hisa Arima},
  \citenamefont {Kawasaki}, \citenamefont {Tokura}, \citenamefont {hisa Arima},
  \citenamefont {Kawasaki},\ and\ \citenamefont {Tokura}}]{Yu2017advMat}%
  \BibitemOpen
  \bibfield  {author} {\bibinfo {author} {\bibfnamefont {Xiuzhen}\ \bibnamefont
  {Yu}}, \bibinfo {author} {\bibfnamefont {Daisuke}\ \bibnamefont {Morikawa}},
  \bibinfo {author} {\bibfnamefont {Yusuke}\ \bibnamefont {Tokunaga}}, \bibinfo
  {author} {\bibfnamefont {Masashi}\ \bibnamefont {Kubota}}, \bibinfo {author}
  {\bibfnamefont {Takashi}\ \bibnamefont {Kurumaji}}, \bibinfo {author}
  {\bibfnamefont {Hiroshi}\ \bibnamefont {Oike}}, \bibinfo {author}
  {\bibfnamefont {Masao}\ \bibnamefont {Nakamura}}, \bibinfo {author}
  {\bibfnamefont {Fumitaka}\ \bibnamefont {Kagawa}}, \bibinfo {author}
  {\bibfnamefont {Yasujiro}\ \bibnamefont {Taguchi}}, \bibinfo {author}
  {\bibfnamefont {Taka}\ \bibnamefont {hisa Arima}}, \bibinfo {author}
  {\bibfnamefont {Masashi}\ \bibnamefont {Kawasaki}}, \bibinfo {author}
  {\bibfnamefont {Yoshinori}\ \bibnamefont {Tokura}}, \bibinfo {author}
  {\bibfnamefont {Taka}\ \bibnamefont {hisa Arima}}, \bibinfo {author}
  {\bibfnamefont {Masashi}\ \bibnamefont {Kawasaki}}, \bibinfo {author}
  {\bibfnamefont {Yoshinori}\ \bibnamefont {Tokura}}, \bibinfo {author}
  {\bibfnamefont {Taka}\ \bibnamefont {hisa Arima}}, \bibinfo {author}
  {\bibfnamefont {Masashi}\ \bibnamefont {Kawasaki}}, \ and\ \bibinfo {author}
  {\bibfnamefont {Yoshinori}\ \bibnamefont {Tokura}},\ }\bibfield  {title}
  {\enquote {\bibinfo {title} {{Current-Induced Nucleation and Annihilation of
  Magnetic Skyrmions at Room Temperature in a Chiral Magnet}},}\ }\href
  {\doibase 10.1002/adma.201606178} {\bibfield  {journal} {\bibinfo  {journal}
  {Advanced Materials}\ } (\bibinfo {year} {2017}),\
  10.1002/adma.201606178}\BibitemShut {NoStop}%
\bibitem [{\citenamefont {Yu}\ \emph {et~al.}(2012)\citenamefont {Yu},
  \citenamefont {Kanazawa}, \citenamefont {Zhang}, \citenamefont {Nagai},
  \citenamefont {Hara}, \citenamefont {Kimoto}, \citenamefont {Matsui},
  \citenamefont {Onose},\ and\ \citenamefont {Tokura}}]{Yu2012NatComFeGe_Jc}%
  \BibitemOpen
  \bibfield  {author} {\bibinfo {author} {\bibfnamefont {X.Z.}\ \bibnamefont
  {Yu}}, \bibinfo {author} {\bibfnamefont {N.}~\bibnamefont {Kanazawa}},
  \bibinfo {author} {\bibfnamefont {W.Z.}\ \bibnamefont {Zhang}}, \bibinfo
  {author} {\bibfnamefont {T.}~\bibnamefont {Nagai}}, \bibinfo {author}
  {\bibfnamefont {T.}~\bibnamefont {Hara}}, \bibinfo {author} {\bibfnamefont
  {K.}~\bibnamefont {Kimoto}}, \bibinfo {author} {\bibfnamefont
  {Y.}~\bibnamefont {Matsui}}, \bibinfo {author} {\bibfnamefont
  {Y.}~\bibnamefont {Onose}}, \ and\ \bibinfo {author} {\bibfnamefont
  {Y.}~\bibnamefont {Tokura}},\ }\bibfield  {title} {\enquote {\bibinfo {title}
  {{Skyrmion flow near room temperature in an ultralow current density}},}\
  }\href {\doibase 10.1038/ncomms1990} {\bibfield  {journal} {\bibinfo
  {journal} {Nature Communications}\ }\textbf {\bibinfo {volume} {3}},\
  \bibinfo {pages} {988} (\bibinfo {year} {2012})}\BibitemShut {NoStop}%
\bibitem [{\citenamefont {McGrouther}\ \emph {et~al.}(2016)\citenamefont
  {McGrouther}, \citenamefont {Lamb}, \citenamefont {Krajnak}, \citenamefont
  {McFadzean}, \citenamefont {McVitie}, \citenamefont {Stamps}, \citenamefont
  {Leonov}, \citenamefont {Bogdanov},\ and\ \citenamefont
  {Togawa}}]{McGrouther2016NJpP_TEM_FeGe}%
  \BibitemOpen
  \bibfield  {author} {\bibinfo {author} {\bibfnamefont {D}~\bibnamefont
  {McGrouther}}, \bibinfo {author} {\bibfnamefont {R~J}\ \bibnamefont {Lamb}},
  \bibinfo {author} {\bibfnamefont {M}~\bibnamefont {Krajnak}}, \bibinfo
  {author} {\bibfnamefont {S}~\bibnamefont {McFadzean}}, \bibinfo {author}
  {\bibfnamefont {S}~\bibnamefont {McVitie}}, \bibinfo {author} {\bibfnamefont
  {R~L}\ \bibnamefont {Stamps}}, \bibinfo {author} {\bibfnamefont {A~O}\
  \bibnamefont {Leonov}}, \bibinfo {author} {\bibfnamefont {A~N}\ \bibnamefont
  {Bogdanov}}, \ and\ \bibinfo {author} {\bibfnamefont {Y}~\bibnamefont
  {Togawa}},\ }\bibfield  {title} {\enquote {\bibinfo {title} {{Internal
  structure of hexagonal skyrmion lattices in cubic helimagnets}},}\ }\href
  {\doibase 10.1088/1367-2630/18/9/095004} {\bibfield  {journal} {\bibinfo
  {journal} {New Journal of Physics}\ }\textbf {\bibinfo {volume} {18}}
  (\bibinfo {year} {2016}),\ 10.1088/1367-2630/18/9/095004}\BibitemShut
  {NoStop}%
\bibitem [{\citenamefont {Stolt}\ \emph {et~al.}(2016)\citenamefont {Stolt},
  \citenamefont {Li}, \citenamefont {Phillips}, \citenamefont {Song},
  \citenamefont {Mathur}, \citenamefont {Dunin-Borkowski},\ and\ \citenamefont
  {Jin}}]{StoltJin2016nanolet_FeGeNW}%
  \BibitemOpen
  \bibfield  {author} {\bibinfo {author} {\bibfnamefont {Matthew~J.}\
  \bibnamefont {Stolt}}, \bibinfo {author} {\bibfnamefont {Zi-An}\ \bibnamefont
  {Li}}, \bibinfo {author} {\bibfnamefont {Brandon}\ \bibnamefont {Phillips}},
  \bibinfo {author} {\bibfnamefont {Dongsheng}\ \bibnamefont {Song}}, \bibinfo
  {author} {\bibfnamefont {Nitish}\ \bibnamefont {Mathur}}, \bibinfo {author}
  {\bibfnamefont {Rafal~E.}\ \bibnamefont {Dunin-Borkowski}}, \ and\ \bibinfo
  {author} {\bibfnamefont {Song}\ \bibnamefont {Jin}},\ }\bibfield  {title}
  {\enquote {\bibinfo {title} {{Selective Chemical Vapor Deposition Growth of
  Cubic FeGe Nanowires That Support Stabilized Magnetic Skyrmions}},}\ }\href
  {\doibase 10.1021/acs.nanolett.6b04548} {\bibfield  {journal} {\bibinfo
  {journal} {Nano Letters}\ ,\ \bibinfo {pages} {acs.nanolett.6b04548}}
  (\bibinfo {year} {2016})}\BibitemShut {NoStop}%
\bibitem [{\citenamefont {Moskvin}\ \emph {et~al.}(2013)\citenamefont
  {Moskvin}, \citenamefont {Grigoriev}, \citenamefont {Dyadkin}, \citenamefont
  {Eckerlebe}, \citenamefont {Baenitz}, \citenamefont {Schmidt},\ and\
  \citenamefont {Wilhelm}}]{Moskvin2013PRLSANSFeGe}%
  \BibitemOpen
  \bibfield  {author} {\bibinfo {author} {\bibfnamefont {E.}~\bibnamefont
  {Moskvin}}, \bibinfo {author} {\bibfnamefont {S.}~\bibnamefont {Grigoriev}},
  \bibinfo {author} {\bibfnamefont {V.}~\bibnamefont {Dyadkin}}, \bibinfo
  {author} {\bibfnamefont {H.}~\bibnamefont {Eckerlebe}}, \bibinfo {author}
  {\bibfnamefont {M.}~\bibnamefont {Baenitz}}, \bibinfo {author} {\bibfnamefont
  {M.}~\bibnamefont {Schmidt}}, \ and\ \bibinfo {author} {\bibfnamefont
  {H.}~\bibnamefont {Wilhelm}},\ }\bibfield  {title} {\enquote {\bibinfo
  {title} {{Complex Chiral Modulations in FeGe Close to Magnetic Ordering}},}\
  }\href {\doibase 10.1103/PhysRevLett.110.077207} {\bibfield  {journal}
  {\bibinfo  {journal} {Phys. Rev. Lett.}\ }\textbf {\bibinfo {volume} {110}},\
  \bibinfo {pages} {077207} (\bibinfo {year} {2013})}\BibitemShut {NoStop}%
\bibitem [{\citenamefont {Wilhelm}\ \emph {et~al.}(2011)\citenamefont
  {Wilhelm}, \citenamefont {Baenitz}, \citenamefont {Schmidt}, \citenamefont
  {R\"o\ss{}ler}, \citenamefont {Leonov},\ and\ \citenamefont
  {Bogdanov}}]{Wilhelm2011PRLFeGemultiphase}%
  \BibitemOpen
  \bibfield  {author} {\bibinfo {author} {\bibfnamefont {H.}~\bibnamefont
  {Wilhelm}}, \bibinfo {author} {\bibfnamefont {M.}~\bibnamefont {Baenitz}},
  \bibinfo {author} {\bibfnamefont {M.}~\bibnamefont {Schmidt}}, \bibinfo
  {author} {\bibfnamefont {U.~K.}\ \bibnamefont {R\"o\ss{}ler}}, \bibinfo
  {author} {\bibfnamefont {A.~A.}\ \bibnamefont {Leonov}}, \ and\ \bibinfo
  {author} {\bibfnamefont {A.~N.}\ \bibnamefont {Bogdanov}},\ }\bibfield
  {title} {\enquote {\bibinfo {title} {{Precursor Phenomena at the Magnetic
  Ordering of the Cubic Helimagnet FeGe}},}\ }\href {\doibase
  10.1103/PhysRevLett.107.127203} {\bibfield  {journal} {\bibinfo  {journal}
  {Phys. Rev. Lett.}\ }\textbf {\bibinfo {volume} {107}},\ \bibinfo {pages}
  {127203} (\bibinfo {year} {2011})}\BibitemShut {NoStop}%
\bibitem [{\citenamefont {Cevey}\ \emph {et~al.}(2013)\citenamefont {Cevey},
  \citenamefont {Wilhelm}, \citenamefont {Schmidt},\ and\ \citenamefont
  {Lortz}}]{CeveyPSS2013FeGe}%
  \BibitemOpen
  \bibfield  {author} {\bibinfo {author} {\bibfnamefont {L.}~\bibnamefont
  {Cevey}}, \bibinfo {author} {\bibfnamefont {H.}~\bibnamefont {Wilhelm}},
  \bibinfo {author} {\bibfnamefont {M.}~\bibnamefont {Schmidt}}, \ and\
  \bibinfo {author} {\bibfnamefont {R.}~\bibnamefont {Lortz}},\ }\bibfield
  {title} {\enquote {\bibinfo {title} {{Thermodynamic investigations in the
  precursor region of FeGe}},}\ }\href {\doibase 10.1002/pssb.201200632}
  {\bibfield  {journal} {\bibinfo  {journal} {Physica Status Solidi (B) Basic
  Research}\ }\textbf {\bibinfo {volume} {250}},\ \bibinfo {pages} {650--653}
  (\bibinfo {year} {2013})}\BibitemShut {NoStop}%
\bibitem [{\citenamefont {Turgut}\ \emph {et~al.}(2017)\citenamefont {Turgut},
  \citenamefont {Stolt}, \citenamefont {Jin},\ and\ \citenamefont
  {Fuchs}}]{StoltJin2017}%
  \BibitemOpen
  \bibfield  {author} {\bibinfo {author} {\bibfnamefont {Emrah}\ \bibnamefont
  {Turgut}}, \bibinfo {author} {\bibfnamefont {Matthew~J}\ \bibnamefont
  {Stolt}}, \bibinfo {author} {\bibfnamefont {Song}\ \bibnamefont {Jin}}, \
  and\ \bibinfo {author} {\bibfnamefont {Gregory~D}\ \bibnamefont {Fuchs}},\
  }\bibfield  {title} {\enquote {\bibinfo {title} {{Topological spin dynamics
  in cubic FeGe near room temperature}},}\ }\href
  {http://arxiv.org/abs/1705.03397} {\  (\bibinfo {year} {2017})},\ \Eprint
  {http://arxiv.org/abs/1705.03397} {arXiv:1705.03397} \BibitemShut {NoStop}%
\bibitem [{\citenamefont {Huang}\ and\ \citenamefont
  {Chien}(2012)}]{Huang2012PRL_FeGe_rho_film}%
  \BibitemOpen
  \bibfield  {author} {\bibinfo {author} {\bibfnamefont {S.~X.}\ \bibnamefont
  {Huang}}\ and\ \bibinfo {author} {\bibfnamefont {C.~L.}\ \bibnamefont
  {Chien}},\ }\bibfield  {title} {\enquote {\bibinfo {title} {{Extended
  Skyrmion Phase in Epitaxial $\mathrm{FeGe}(111)$ Thin Films}},}\ }\href
  {\doibase 10.1103/PhysRevLett.108.267201} {\bibfield  {journal} {\bibinfo
  {journal} {Phys. Rev. Lett.}\ }\textbf {\bibinfo {volume} {108}},\ \bibinfo
  {pages} {267201} (\bibinfo {year} {2012})}\BibitemShut {NoStop}%
\bibitem [{\citenamefont {Porter}\ \emph {et~al.}(2014)\citenamefont {Porter},
  \citenamefont {Gartside},\ and\ \citenamefont
  {Marrows}}]{Porter2014PRBFeGefilmTransport}%
  \BibitemOpen
  \bibfield  {author} {\bibinfo {author} {\bibfnamefont {N.~A.}\ \bibnamefont
  {Porter}}, \bibinfo {author} {\bibfnamefont {J.~C.}\ \bibnamefont
  {Gartside}}, \ and\ \bibinfo {author} {\bibfnamefont {C.~H.}\ \bibnamefont
  {Marrows}},\ }\bibfield  {title} {\enquote {\bibinfo {title} {{Scattering
  mechanisms in textured FeGe thin films: Magnetoresistance and the anomalous
  Hall effect}},}\ }\href {\doibase 10.1103/PhysRevB.90.024403} {\bibfield
  {journal} {\bibinfo  {journal} {Phys. Rev. B}\ }\textbf {\bibinfo {volume}
  {90}},\ \bibinfo {pages} {024403} (\bibinfo {year} {2014})}\BibitemShut
  {NoStop}%
\bibitem [{\citenamefont {Kanazawa}\ \emph {et~al.}(2015)\citenamefont
  {Kanazawa}, \citenamefont {Kubota}, \citenamefont {Tsukazaki}, \citenamefont
  {Kozuka}, \citenamefont {Takahashi}, \citenamefont {Kawasaki}, \citenamefont
  {Ichikawa}, \citenamefont {Kagawa},\ and\ \citenamefont
  {Tokura}}]{Kanazawa2015PRB_FeGeconstricted}%
  \BibitemOpen
  \bibfield  {author} {\bibinfo {author} {\bibfnamefont {N.}~\bibnamefont
  {Kanazawa}}, \bibinfo {author} {\bibfnamefont {M.}~\bibnamefont {Kubota}},
  \bibinfo {author} {\bibfnamefont {A.}~\bibnamefont {Tsukazaki}}, \bibinfo
  {author} {\bibfnamefont {Y.}~\bibnamefont {Kozuka}}, \bibinfo {author}
  {\bibfnamefont {K.~S.}\ \bibnamefont {Takahashi}}, \bibinfo {author}
  {\bibfnamefont {M.}~\bibnamefont {Kawasaki}}, \bibinfo {author}
  {\bibfnamefont {M.}~\bibnamefont {Ichikawa}}, \bibinfo {author}
  {\bibfnamefont {F.}~\bibnamefont {Kagawa}}, \ and\ \bibinfo {author}
  {\bibfnamefont {Y.}~\bibnamefont {Tokura}},\ }\bibfield  {title} {\enquote
  {\bibinfo {title} {{Discretized topological Hall effect emerging from
  skyrmions in constricted geometry}},}\ }\href {\doibase
  10.1103/PhysRevB.91.041122} {\bibfield  {journal} {\bibinfo  {journal} {Phys.
  Rev. B}\ }\textbf {\bibinfo {volume} {91}},\ \bibinfo {pages} {041122}
  (\bibinfo {year} {2015})}\BibitemShut {NoStop}%
\bibitem [{\citenamefont {Zhang}\ \emph {et~al.}(2017)\citenamefont {Zhang},
  \citenamefont {Stasinopoulos}, \citenamefont {Lancaster}, \citenamefont
  {Xiao}, \citenamefont {Bauer}, \citenamefont {Rucker}, \citenamefont {Baker},
  \citenamefont {Figueroa}, \citenamefont {Salman}, \citenamefont {Pratt},
  \citenamefont {Blundell}, \citenamefont {Prokscha}, \citenamefont {Suter},
  \citenamefont {Waizner}, \citenamefont {Garst}, \citenamefont {Grundler},
  \citenamefont {van~der Laan}, \citenamefont {Pfleiderer},\ and\ \citenamefont
  {Hesjedal}}]{Zhang2017SciRepFeGefilmswithoutSkX}%
  \BibitemOpen
  \bibfield  {author} {\bibinfo {author} {\bibfnamefont {S~L}\ \bibnamefont
  {Zhang}}, \bibinfo {author} {\bibfnamefont {I}~\bibnamefont {Stasinopoulos}},
  \bibinfo {author} {\bibfnamefont {T}~\bibnamefont {Lancaster}}, \bibinfo
  {author} {\bibfnamefont {F}~\bibnamefont {Xiao}}, \bibinfo {author}
  {\bibfnamefont {A}~\bibnamefont {Bauer}}, \bibinfo {author} {\bibfnamefont
  {F}~\bibnamefont {Rucker}}, \bibinfo {author} {\bibfnamefont {A~A}\
  \bibnamefont {Baker}}, \bibinfo {author} {\bibfnamefont {A.~I.}\ \bibnamefont
  {Figueroa}}, \bibinfo {author} {\bibfnamefont {Z.}~\bibnamefont {Salman}},
  \bibinfo {author} {\bibfnamefont {F.~L.}\ \bibnamefont {Pratt}}, \bibinfo
  {author} {\bibfnamefont {S.~J.}\ \bibnamefont {Blundell}}, \bibinfo {author}
  {\bibfnamefont {T.}~\bibnamefont {Prokscha}}, \bibinfo {author}
  {\bibfnamefont {A.}~\bibnamefont {Suter}}, \bibinfo {author} {\bibfnamefont
  {J.}~\bibnamefont {Waizner}}, \bibinfo {author} {\bibfnamefont
  {M.}~\bibnamefont {Garst}}, \bibinfo {author} {\bibfnamefont
  {D.}~\bibnamefont {Grundler}}, \bibinfo {author} {\bibfnamefont
  {G.}~\bibnamefont {van~der Laan}}, \bibinfo {author} {\bibfnamefont
  {C.}~\bibnamefont {Pfleiderer}}, \ and\ \bibinfo {author} {\bibfnamefont
  {T.}~\bibnamefont {Hesjedal}},\ }\bibfield  {title} {\enquote {\bibinfo
  {title} {{Room-temperature helimagnetism in FeGe thin films}},}\ }\href
  {\doibase 10.1038/s41598-017-00201-z} {\bibfield  {journal} {\bibinfo
  {journal} {Scientific Reports}\ }\textbf {\bibinfo {volume} {7}},\ \bibinfo
  {pages} {123} (\bibinfo {year} {2017})}\BibitemShut {NoStop}%
\bibitem [{\citenamefont {Meynell}\ \emph {et~al.}(2014)\citenamefont
  {Meynell}, \citenamefont {Wilson}, \citenamefont {Loudon}, \citenamefont
  {Spitzig}, \citenamefont {Rybakov}, \citenamefont {Johnson},\ and\
  \citenamefont {Monchesky}}]{MoncheskyPRBquestionFeGethinfilms}%
  \BibitemOpen
  \bibfield  {author} {\bibinfo {author} {\bibfnamefont {S.~A.}\ \bibnamefont
  {Meynell}}, \bibinfo {author} {\bibfnamefont {M.~N.}\ \bibnamefont {Wilson}},
  \bibinfo {author} {\bibfnamefont {J.~C.}\ \bibnamefont {Loudon}}, \bibinfo
  {author} {\bibfnamefont {A.}~\bibnamefont {Spitzig}}, \bibinfo {author}
  {\bibfnamefont {F.~N.}\ \bibnamefont {Rybakov}}, \bibinfo {author}
  {\bibfnamefont {M.~B.}\ \bibnamefont {Johnson}}, \ and\ \bibinfo {author}
  {\bibfnamefont {T.~L.}\ \bibnamefont {Monchesky}},\ }\bibfield  {title}
  {\enquote {\bibinfo {title} {{Hall effect and transmission electron
  microscopy of epitaxial MnSi thin films}},}\ }\href {\doibase
  10.1103/PhysRevB.90.224419} {\bibfield  {journal} {\bibinfo  {journal} {Phys.
  Rev. B}\ }\textbf {\bibinfo {volume} {90}},\ \bibinfo {pages} {224419}
  (\bibinfo {year} {2014})}\BibitemShut {NoStop}%
\bibitem [{\citenamefont {Monchesky}\ \emph {et~al.}(2014)\citenamefont
  {Monchesky}, \citenamefont {Loudon}, \citenamefont {Robertson},\ and\
  \citenamefont {Bogdanov}}]{MoncheskyCommenttoPRL}%
  \BibitemOpen
  \bibfield  {author} {\bibinfo {author} {\bibfnamefont {T.~L.}\ \bibnamefont
  {Monchesky}}, \bibinfo {author} {\bibfnamefont {J.~C.}\ \bibnamefont
  {Loudon}}, \bibinfo {author} {\bibfnamefont {M.~D.}\ \bibnamefont
  {Robertson}}, \ and\ \bibinfo {author} {\bibfnamefont {A.~N.}\ \bibnamefont
  {Bogdanov}},\ }\bibfield  {title} {\enquote {\bibinfo {title} {{Comment on
  ``Robust Formation of Skyrmions and Topological Hall Effect Anomaly in
  Epitaxial Thin Films of MnSi''}},}\ }\href {\doibase
  10.1103/PhysRevLett.112.059701} {\bibfield  {journal} {\bibinfo  {journal}
  {Phys. Rev. Lett.}\ }\textbf {\bibinfo {volume} {112}},\ \bibinfo {pages}
  {059701} (\bibinfo {year} {2014})}\BibitemShut {NoStop}%
\bibitem [{\citenamefont {Wilhelm}\ \emph {et~al.}(2016)\citenamefont
  {Wilhelm}, \citenamefont {Leonov}, \citenamefont {R\"o\ss{}ler},
  \citenamefont {Burger}, \citenamefont {Hardy}, \citenamefont {Meingast},
  \citenamefont {Gruner}, \citenamefont {Schnelle}, \citenamefont {Schmidt},\
  and\ \citenamefont {Baenitz}}]{Wilhelm2016PRB_FeGescaling}%
  \BibitemOpen
  \bibfield  {author} {\bibinfo {author} {\bibfnamefont {H.}~\bibnamefont
  {Wilhelm}}, \bibinfo {author} {\bibfnamefont {A.~O.}\ \bibnamefont {Leonov}},
  \bibinfo {author} {\bibfnamefont {U.~K.}\ \bibnamefont {R\"o\ss{}ler}},
  \bibinfo {author} {\bibfnamefont {P.}~\bibnamefont {Burger}}, \bibinfo
  {author} {\bibfnamefont {F.}~\bibnamefont {Hardy}}, \bibinfo {author}
  {\bibfnamefont {C.}~\bibnamefont {Meingast}}, \bibinfo {author}
  {\bibfnamefont {M.~E.}\ \bibnamefont {Gruner}}, \bibinfo {author}
  {\bibfnamefont {W.}~\bibnamefont {Schnelle}}, \bibinfo {author}
  {\bibfnamefont {M.}~\bibnamefont {Schmidt}}, \ and\ \bibinfo {author}
  {\bibfnamefont {M.}~\bibnamefont {Baenitz}},\ }\bibfield  {title} {\enquote
  {\bibinfo {title} {{Scaling study and thermodynamic properties of the cubic
  helimagnet FeGe}},}\ }\href {\doibase 10.1103/PhysRevB.94.144424} {\bibfield
  {journal} {\bibinfo  {journal} {Phys. Rev. B}\ }\textbf {\bibinfo {volume}
  {94}},\ \bibinfo {pages} {144424} (\bibinfo {year} {2016})}\BibitemShut
  {NoStop}%
\bibitem [{\citenamefont {Yeo}\ \emph {et~al.}(2003)\citenamefont {Yeo},
  \citenamefont {Nakatsuji}, \citenamefont {Bianchi}, \citenamefont
  {Schlottmann}, \citenamefont {Fisk}, \citenamefont {Balicas}, \citenamefont
  {Stampe},\ and\ \citenamefont {Kennedy}}]{Yeo2003PRLFeSiGe}%
  \BibitemOpen
  \bibfield  {author} {\bibinfo {author} {\bibfnamefont {S.}~\bibnamefont
  {Yeo}}, \bibinfo {author} {\bibfnamefont {S.}~\bibnamefont {Nakatsuji}},
  \bibinfo {author} {\bibfnamefont {A.~D.}\ \bibnamefont {Bianchi}}, \bibinfo
  {author} {\bibfnamefont {P.}~\bibnamefont {Schlottmann}}, \bibinfo {author}
  {\bibfnamefont {Z.}~\bibnamefont {Fisk}}, \bibinfo {author} {\bibfnamefont
  {L.}~\bibnamefont {Balicas}}, \bibinfo {author} {\bibfnamefont {P.~A.}\
  \bibnamefont {Stampe}}, \ and\ \bibinfo {author} {\bibfnamefont {R.~J.}\
  \bibnamefont {Kennedy}},\ }\bibfield  {title} {\enquote {\bibinfo {title}
  {{First-Order Transition from a Kondo Insulator to a Ferromagnetic Metal in
  Single Crystalline FeSi$_{1-x}$Ge$_x$}},}\ }\href {\doibase
  10.1103/PhysRevLett.91.046401} {\bibfield  {journal} {\bibinfo  {journal}
  {Physical Review Letters}\ }\textbf {\bibinfo {volume} {91}},\ \bibinfo
  {pages} {046401} (\bibinfo {year} {2003})}\BibitemShut {NoStop}%
\bibitem [{\citenamefont {Pedrazzini}\ \emph {et~al.}(2007)\citenamefont
  {Pedrazzini}, \citenamefont {Wilhelm}, \citenamefont {Jaccard}, \citenamefont
  {Jarlborg}, \citenamefont {Schmidt}, \citenamefont {Hanfland}, \citenamefont
  {Akselrud}, \citenamefont {Yuan}, \citenamefont {Schwarz}, \citenamefont
  {Grin},\ and\ \citenamefont {Steglich}}]{Pedrazzini2007PRLFeGe_rhoxx_vs_P}%
  \BibitemOpen
  \bibfield  {author} {\bibinfo {author} {\bibfnamefont {P.}~\bibnamefont
  {Pedrazzini}}, \bibinfo {author} {\bibfnamefont {H.}~\bibnamefont {Wilhelm}},
  \bibinfo {author} {\bibfnamefont {D.}~\bibnamefont {Jaccard}}, \bibinfo
  {author} {\bibfnamefont {T.}~\bibnamefont {Jarlborg}}, \bibinfo {author}
  {\bibfnamefont {M.}~\bibnamefont {Schmidt}}, \bibinfo {author} {\bibfnamefont
  {M.}~\bibnamefont {Hanfland}}, \bibinfo {author} {\bibfnamefont
  {L.}~\bibnamefont {Akselrud}}, \bibinfo {author} {\bibfnamefont {H.~Q.}\
  \bibnamefont {Yuan}}, \bibinfo {author} {\bibfnamefont {U.}~\bibnamefont
  {Schwarz}}, \bibinfo {author} {\bibfnamefont {Yu.}\ \bibnamefont {Grin}}, \
  and\ \bibinfo {author} {\bibfnamefont {F.}~\bibnamefont {Steglich}},\
  }\bibfield  {title} {\enquote {\bibinfo {title} {{Metallic State in Cubic
  FeGe Beyond Its Quantum Phase Transition}},}\ }\href {\doibase
  10.1103/PhysRevLett.98.047204} {\bibfield  {journal} {\bibinfo  {journal}
  {Phys. Rev. Lett.}\ }\textbf {\bibinfo {volume} {98}},\ \bibinfo {pages}
  {047204} (\bibinfo {year} {2007})}\BibitemShut {NoStop}%
\bibitem [{\citenamefont {Neubauer}\ \emph
  {et~al.}(2009{\natexlab{b}})\citenamefont {Neubauer}, \citenamefont
  {Pfleiderer}, \citenamefont {Binz}, \citenamefont {Rosch}, \citenamefont
  {Ritz}, \citenamefont {Niklowitz},\ and\ \citenamefont
  {B\"oni}}]{Neubauer2009PRL_THE_MnSi}%
  \BibitemOpen
  \bibfield  {author} {\bibinfo {author} {\bibfnamefont {A.}~\bibnamefont
  {Neubauer}}, \bibinfo {author} {\bibfnamefont {C.}~\bibnamefont
  {Pfleiderer}}, \bibinfo {author} {\bibfnamefont {B.}~\bibnamefont {Binz}},
  \bibinfo {author} {\bibfnamefont {A.}~\bibnamefont {Rosch}}, \bibinfo
  {author} {\bibfnamefont {R.}~\bibnamefont {Ritz}}, \bibinfo {author}
  {\bibfnamefont {P.~G.}\ \bibnamefont {Niklowitz}}, \ and\ \bibinfo {author}
  {\bibfnamefont {P.}~\bibnamefont {B\"oni}},\ }\bibfield  {title} {\enquote
  {\bibinfo {title} {{Topological Hall Effect in the $A$ Phase of MnSi}},}\
  }\href {\doibase 10.1103/PhysRevLett.102.186602} {\bibfield  {journal}
  {\bibinfo  {journal} {Phys. Rev. Lett.}\ }\textbf {\bibinfo {volume} {102}},\
  \bibinfo {pages} {186602} (\bibinfo {year} {2009}{\natexlab{b}})}\BibitemShut
  {NoStop}%
\bibitem [{\citenamefont {Ritz}(2010)}]{Ritz2010PhDThesis}%
  \BibitemOpen
  \bibfield  {author} {\bibinfo {author} {\bibfnamefont {R.}~\bibnamefont
  {Ritz}},\ }in\ \href {http://mediatum.ub.tum.de/doc/1097668/1097668.pdf}
  {\emph {\bibinfo {booktitle} {Ph.D. Thesis, Superconductivity and non-Fermi
  liquid behavior on the border of itinerant ferromagnetism}}}\ (\bibinfo
  {year} {2010})\ pp.\ \bibinfo {pages} {113--115}\BibitemShut {NoStop}%
\bibitem [{\citenamefont {Ritz}\ \emph {et~al.}(2013)\citenamefont {Ritz},
  \citenamefont {Halder}, \citenamefont {Franz}, \citenamefont {Bauer},
  \citenamefont {Wagner}, \citenamefont {Bamler}, \citenamefont {Rosch},\ and\
  \citenamefont {Pfleiderer}}]{Ritz2013PRB_THE_pressure}%
  \BibitemOpen
  \bibfield  {author} {\bibinfo {author} {\bibfnamefont {R.}~\bibnamefont
  {Ritz}}, \bibinfo {author} {\bibfnamefont {M.}~\bibnamefont {Halder}},
  \bibinfo {author} {\bibfnamefont {C.}~\bibnamefont {Franz}}, \bibinfo
  {author} {\bibfnamefont {A.}~\bibnamefont {Bauer}}, \bibinfo {author}
  {\bibfnamefont {M.}~\bibnamefont {Wagner}}, \bibinfo {author} {\bibfnamefont
  {R.}~\bibnamefont {Bamler}}, \bibinfo {author} {\bibfnamefont
  {A.}~\bibnamefont {Rosch}}, \ and\ \bibinfo {author} {\bibfnamefont
  {C.}~\bibnamefont {Pfleiderer}},\ }\bibfield  {title} {\enquote {\bibinfo
  {title} {{Giant generic topological Hall resistivity of MnSi under
  pressure}},}\ }\href {\doibase 10.1103/PhysRevB.87.134424} {\bibfield
  {journal} {\bibinfo  {journal} {Phys. Rev. B}\ }\textbf {\bibinfo {volume}
  {87}},\ \bibinfo {pages} {134424} (\bibinfo {year} {2013})}\BibitemShut
  {NoStop}%
\bibitem [{\citenamefont {Everschor-Sitte}\ and\ \citenamefont
  {Sitte}(2014)}]{EverschorBerryphaseTutorial}%
  \BibitemOpen
  \bibfield  {author} {\bibinfo {author} {\bibfnamefont {Karin}\ \bibnamefont
  {Everschor-Sitte}}\ and\ \bibinfo {author} {\bibfnamefont {Matthias}\
  \bibnamefont {Sitte}},\ }\bibfield  {title} {\enquote {\bibinfo {title}
  {{Real-space Berry phases: Skyrmion soccer (invited)}},}\ }\href {\doibase
  10.1063/1.4870695} {\bibfield  {journal} {\bibinfo  {journal} {Journal of
  Applied Physics}\ }\textbf {\bibinfo {volume} {115}},\ \bibinfo {pages}
  {172602} (\bibinfo {year} {2014})}\BibitemShut {NoStop}%
\bibitem [{\citenamefont {Denisov}\ \emph {et~al.}(2017)\citenamefont
  {Denisov}, \citenamefont {Rozhansky}, \citenamefont {Averkiev},\ and\
  \citenamefont {L{\"{a}}hderanta}}]{Denisov2017}%
  \BibitemOpen
  \bibfield  {author} {\bibinfo {author} {\bibfnamefont {K.~S.}\ \bibnamefont
  {Denisov}}, \bibinfo {author} {\bibfnamefont {I.~V.}\ \bibnamefont
  {Rozhansky}}, \bibinfo {author} {\bibfnamefont {N.~S.}\ \bibnamefont
  {Averkiev}}, \ and\ \bibinfo {author} {\bibfnamefont {E.}~\bibnamefont
  {L{\"{a}}hderanta}},\ }\bibfield  {title} {\enquote {\bibinfo {title} {{A
  nontrivial crossover in topological Hall effect regimes}},}\ }\href {\doibase
  10.1038/s41598-017-16538-4} {\bibfield  {journal} {\bibinfo  {journal}
  {Scientific Reports}\ }\textbf {\bibinfo {volume} {7}},\ \bibinfo {pages}
  {17204} (\bibinfo {year} {2017})}\BibitemShut {NoStop}%
\bibitem [{\citenamefont {Kanazawa}\ \emph {et~al.}(2016)\citenamefont
  {Kanazawa}, \citenamefont {Nii}, \citenamefont {Zhang}, \citenamefont
  {Mishchenko}, \citenamefont {{De Filippis}}, \citenamefont {Kagawa},
  \citenamefont {Iwasa}, \citenamefont {Nagaosa},\ and\ \citenamefont
  {Tokura}}]{Kanazawa2016MnGe}%
  \BibitemOpen
  \bibfield  {author} {\bibinfo {author} {\bibfnamefont {N.}~\bibnamefont
  {Kanazawa}}, \bibinfo {author} {\bibfnamefont {Y.}~\bibnamefont {Nii}},
  \bibinfo {author} {\bibfnamefont {X.~X.}\ \bibnamefont {Zhang}}, \bibinfo
  {author} {\bibfnamefont {A.~S.}\ \bibnamefont {Mishchenko}}, \bibinfo
  {author} {\bibfnamefont {G.}~\bibnamefont {{De Filippis}}}, \bibinfo {author}
  {\bibfnamefont {F.}~\bibnamefont {Kagawa}}, \bibinfo {author} {\bibfnamefont
  {Y.}~\bibnamefont {Iwasa}}, \bibinfo {author} {\bibfnamefont
  {N.}~\bibnamefont {Nagaosa}}, \ and\ \bibinfo {author} {\bibfnamefont
  {Y.}~\bibnamefont {Tokura}},\ }\bibfield  {title} {\enquote {\bibinfo {title}
  {{Critical phenomena of emergent magnetic monopoles in a chiral magnet}},}\
  }\href {\doibase 10.1038/ncomms11622} {\bibfield  {journal} {\bibinfo
  {journal} {Nature Communications}\ }\textbf {\bibinfo {volume} {7}},\
  \bibinfo {pages} {11622} (\bibinfo {year} {2016})}\BibitemShut {NoStop}%
\bibitem [{\citenamefont {Luo}\ \emph {et~al.}(2017)\citenamefont {Luo},
  \citenamefont {Lin}, \citenamefont {Fobes}, \citenamefont {Leroux},
  \citenamefont {Wakeham}, \citenamefont {Bauer}, \citenamefont {Betts},
  \citenamefont {Thompson}, \citenamefont {Miglior}, \citenamefont
  {Janoschek},\ and\ \citenamefont {Maiorov}}]{YK_arxiv}%
  \BibitemOpen
  \bibfield  {author} {\bibinfo {author} {\bibfnamefont {Yongkang}\
  \bibnamefont {Luo}}, \bibinfo {author} {\bibfnamefont {Shizeng}\ \bibnamefont
  {Lin}}, \bibinfo {author} {\bibfnamefont {D.~M.}\ \bibnamefont {Fobes}},
  \bibinfo {author} {\bibfnamefont {M.}~\bibnamefont {Leroux}}, \bibinfo
  {author} {\bibfnamefont {N.}~\bibnamefont {Wakeham}}, \bibinfo {author}
  {\bibfnamefont {E.~D.}\ \bibnamefont {Bauer}}, \bibinfo {author}
  {\bibfnamefont {J.~B.}\ \bibnamefont {Betts}}, \bibinfo {author}
  {\bibfnamefont {J.~D.}\ \bibnamefont {Thompson}}, \bibinfo {author}
  {\bibfnamefont {A.}~\bibnamefont {Miglior}}, \bibinfo {author} {\bibfnamefont
  {M.}~\bibnamefont {Janoschek}}, \ and\ \bibinfo {author} {\bibfnamefont
  {Boris}\ \bibnamefont {Maiorov}},\ }\bibfield  {title} {\enquote {\bibinfo
  {title} {{Depinning currents and elasticity of skyrmion lattice measured via
  Resonant Ultrasound Resonance}},}\ }\href@noop {} {\bibfield  {journal}
  {\bibinfo  {journal} {arXiv}\ }\textbf {\bibinfo {volume} {x}},\ \bibinfo
  {pages} {XXX} (\bibinfo {year} {2017})}\BibitemShut {NoStop}%
\bibitem [{\citenamefont {Richardson}(1967)}]{Richardson1967FeGeCVT}%
  \BibitemOpen
  \bibfield  {author} {\bibinfo {author} {\bibfnamefont {Marcus}\ \bibnamefont
  {Richardson}},\ }\bibfield  {title} {\enquote {\bibinfo {title} {{The Partial
  Equilibrium Diagram of the Fe-Ge System in the Range 40-72 at. \% Ge, and the
  Crystallisation of some Iron Germanides by Chemical Transport Reactions.}}}\
  }\href {\doibase 10.3891/acta.chem.scand.21-2305} {\bibfield  {journal}
  {\bibinfo  {journal} {Acta Chemica Scandinavica}\ }\textbf {\bibinfo {volume}
  {21}},\ \bibinfo {pages} {2305--2317} (\bibinfo {year} {1967})}\BibitemShut
  {NoStop}%
\bibitem [{\citenamefont {Bauer}\ \emph {et~al.}(2010)\citenamefont {Bauer},
  \citenamefont {Neubauer}, \citenamefont {Franz}, \citenamefont
  {M{\"{u}}nzer}, \citenamefont {Garst},\ and\ \citenamefont
  {Pfleiderer}}]{Bauer2010PRBMnFeSiMnCoSi}%
  \BibitemOpen
  \bibfield  {author} {\bibinfo {author} {\bibfnamefont {A.}~\bibnamefont
  {Bauer}}, \bibinfo {author} {\bibfnamefont {A.}~\bibnamefont {Neubauer}},
  \bibinfo {author} {\bibfnamefont {C.}~\bibnamefont {Franz}}, \bibinfo
  {author} {\bibfnamefont {W.}~\bibnamefont {M{\"{u}}nzer}}, \bibinfo {author}
  {\bibfnamefont {M.}~\bibnamefont {Garst}}, \ and\ \bibinfo {author}
  {\bibfnamefont {C.}~\bibnamefont {Pfleiderer}},\ }\bibfield  {title}
  {\enquote {\bibinfo {title} {{Quantum phase transitions in single-crystal
  Mn$_{1-x}$Fe$_x$Si and Mn$_{1-x}$Co$_x$Si: Crystal growth, magnetization, ac
  susceptibility, and specific heat}},}\ }\href {\doibase
  10.1103/PhysRevB.82.064404} {\bibfield  {journal} {\bibinfo  {journal}
  {Physical Review B}\ }\textbf {\bibinfo {volume} {82}},\ \bibinfo {pages}
  {064404} (\bibinfo {year} {2010})}\BibitemShut {NoStop}%
\bibitem [{\citenamefont {Lebech}\ \emph {et~al.}(1989)\citenamefont {Lebech},
  \citenamefont {Bernhard},\ and\ \citenamefont
  {Freltoft}}]{Lebech1989JPCM_FeGe_neutron}%
  \BibitemOpen
  \bibfield  {author} {\bibinfo {author} {\bibfnamefont {B}~\bibnamefont
  {Lebech}}, \bibinfo {author} {\bibfnamefont {J}~\bibnamefont {Bernhard}}, \
  and\ \bibinfo {author} {\bibfnamefont {T}~\bibnamefont {Freltoft}},\
  }\bibfield  {title} {\enquote {\bibinfo {title} {{Magnetic structures of
  cubic FeGe studied by small-angle neutron scattering}},}\ }\href {\doibase
  10.1088/0953-8984/1/35/010} {\bibfield  {journal} {\bibinfo  {journal}
  {Journal of Physics: Condensed Matter}\ }\textbf {\bibinfo {volume} {1}},\
  \bibinfo {pages} {6105--6122} (\bibinfo {year} {1989})}\BibitemShut {NoStop}%
\end{thebibliography}

%

\end{document}


\title{Supplementary Information: Skyrmion Topological Hall Effect near Room Temperature}


\author{Maxime Leroux}
\affiliation{Materials Physics and Applications division, Los Alamos National Laboratory, Los Alamos, New Mexico 87545, United States}

\author{Matthew J. Stolt}
\affiliation{Department of Chemistry, University of Wisconsin-Madison, 1101 University Avenue, Madison, Wisconsin 53706, USA}

\author{Song Jin}
\affiliation{Department of Chemistry, University of Wisconsin-Madison, 1101 University Avenue, Madison, Wisconsin 53706, USA}

\author{Douglas V. Pete}
\affiliation{Center for Integrated Nanotechnologies, Sandia National Laboratories, Albuquerque, New Mexico 87185, United States}

\author{Charles Reichhardt}
\affiliation{Theoretical Division, Los Alamos National Laboratory, Los Alamos, New Mexico 87545, United States}

\author{Boris Maiorov}
\affiliation{Materials Physics and Applications division, Los Alamos National Laboratory, Los Alamos, New Mexico 87545, United States}

\begin{abstract}
In this supplementary information we discuss sample thickness and bulk behavior. We show colormaps made either from the $\rho_{yx}(T)$ curves only (Fig.~\ref{fig1}), or from the $\rho_{yx}(H)$ curves only (Fig.~\ref{fig2}). We also show the full hysteresis cycle measured at 276\,K (Fig.~\ref{fig3}), and finally present the detailed calculations of the demagnetizing factor.
\end{abstract}

\maketitle

\section{Sample Thickness and Bulk Behavior}
Although the lamella sample is thin, several experimental results indicate that it displays bulk behavior.
At 750\,nm thick, our sample is $\approx800$ times thinner than that of Ref.~\cite{Wilhelm2011PRLFeGemultiphase,Wilhelm2016PRB_FeGescaling}, but is still more than 10 times thicker than the helical wavelength\cite{Lebech1989JPCM_FeGe_neutron} $\lambda_S=69.8$\,nm. According to Lorentz TEM\cite{Yu2011NatMat}, a factor of 10 is in the bulk limit in terms of skyrmions phase extension. 
Additionally, the conical to field polarized transition line is in good agreement with the transition line measured in a bulk spherical sample in Ref.~\cite{Wilhelm2011PRLFeGemultiphase,Wilhelm2016PRB_FeGescaling}, after correcting for demagnetizing effects ($B_{int}$ right axis of Fig.~4 of the main text). This transition is defined in Fig.~2.b of the main text and reported as black circles in Fig.~4 of the main text. The high field dependence is also in excellent agreement with the exponent $\beta=0.368$ for 3D-Heisenberg spins, as measured in a previous scaling study of FeGe\cite{Wilhelm2016PRB_FeGescaling}. Finally, the extrapolated value $T_N=278.2$\,K, is very close to  $T_N=278.6$\,K found in Ref.~\cite{Wilhelm2016PRB_FeGescaling}.
As our skyrmion phase extension and the conical to field polarized transition line are both in good quantitative agreement with SANS and magnetization data, we thus observe a H-T phase diagram that is in excellent agreement with that of bulk samples.

\section{HT Diagrams}
\begin{figure}
  \includegraphics[width=0.5\textwidth]{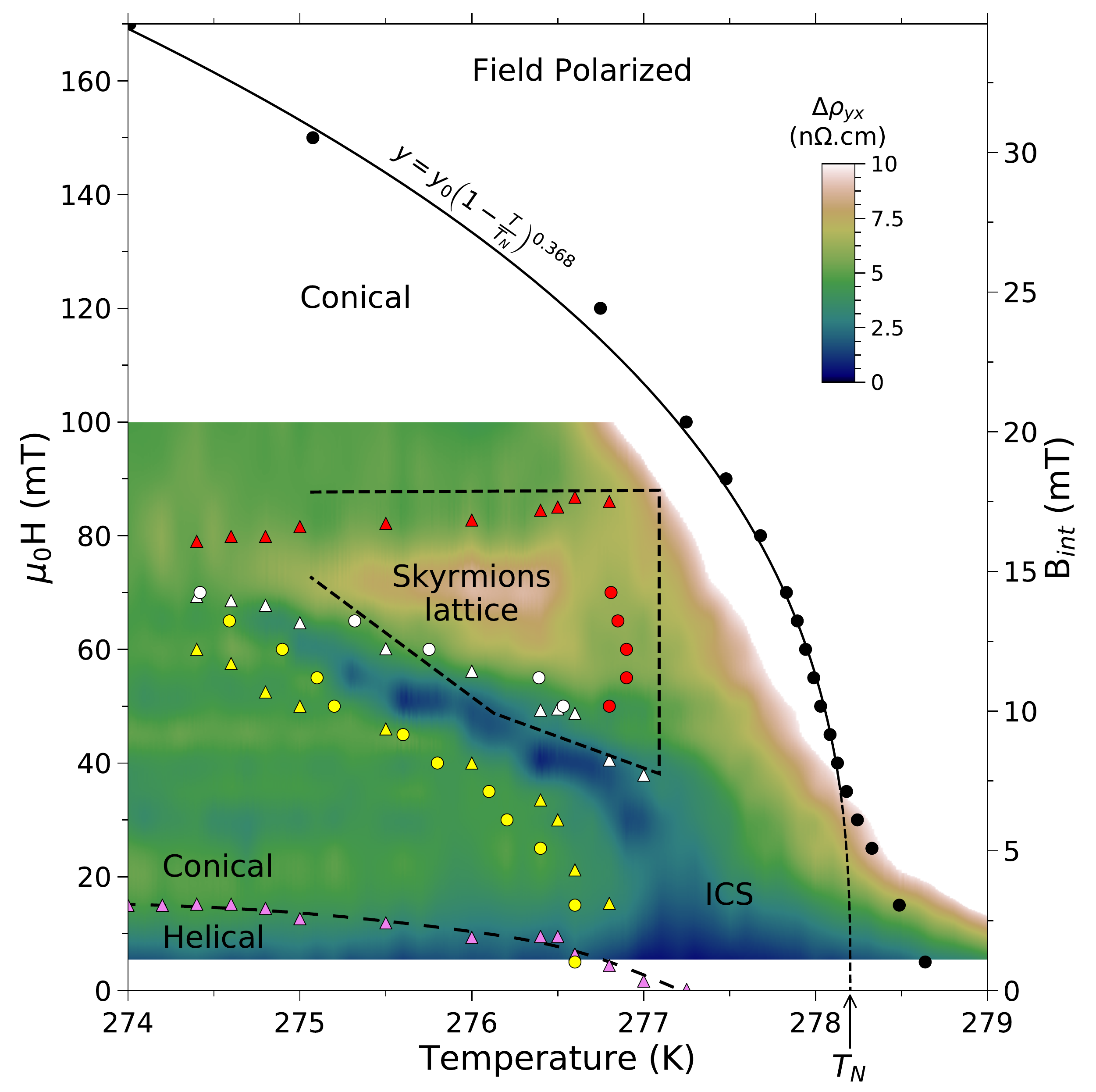}
  \caption{\textbf{HT diagram of the deviation to the linear Hall effect from $\rho_{yx}(T)$ curves only.}
}
  \label{fig1}
\end{figure}

\begin{figure}
  \includegraphics[width=0.5\textwidth]{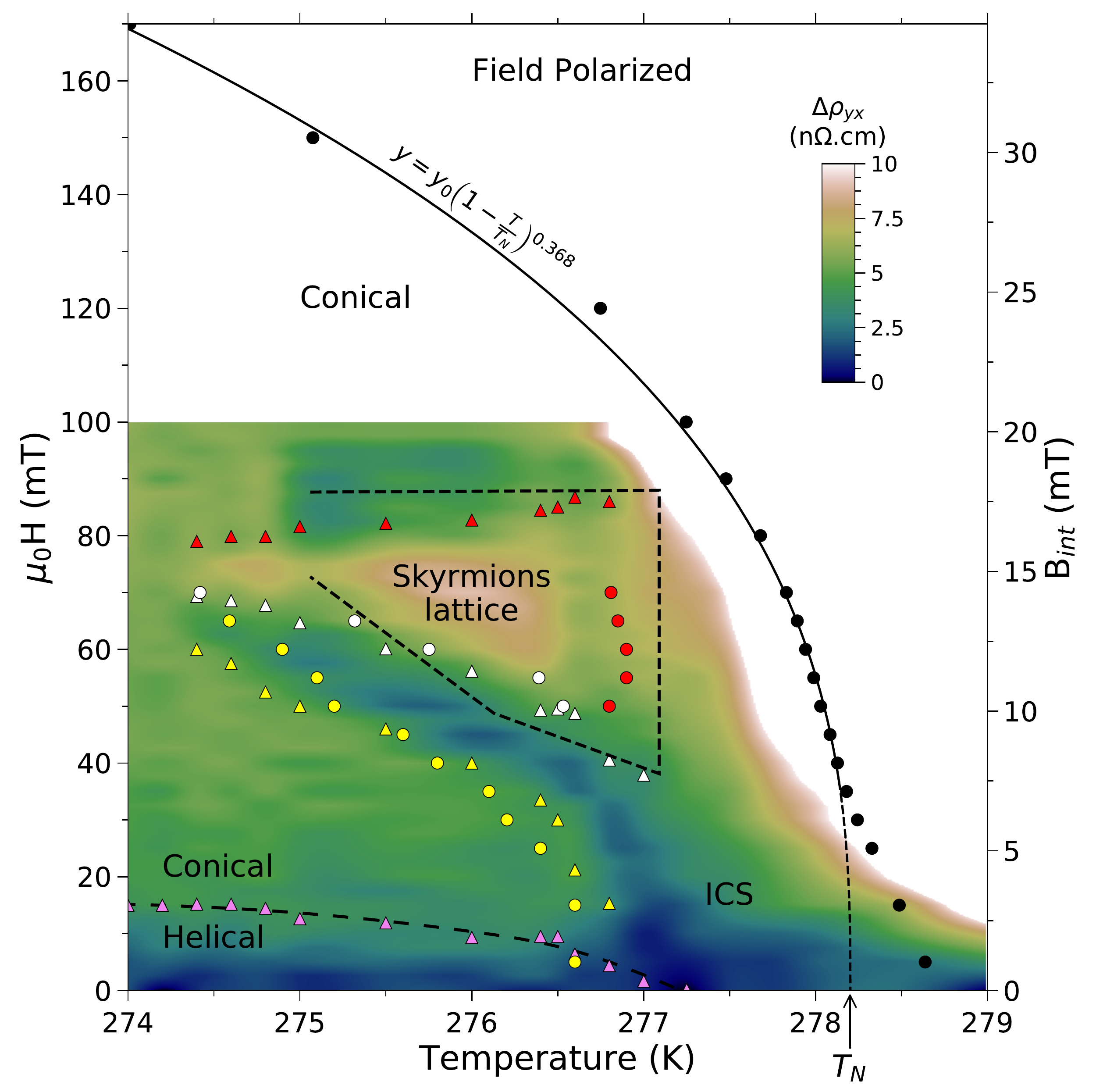}
  \caption{\textbf{HT diagram of the deviation to the linear Hall effect from $\rho_{yx}(H)$ curves only.}  
}
  \label{fig2}
\end{figure}

In Fig.~\ref{fig1} and Fig.~\ref{fig2} we show colormaps made, respectively, either from the $\rho_{yx}(T)$ curves only, or from the $\rho_{yx}(H)$ curves only. 
We recall the legend for the HT diagrams of Fig.~\ref{fig1} and \ref{fig2}. \textit{(Black circles)} transition separating the conical and field polarized phases, as defined by the 5\% criterion in $\rho_{yx}(T)$ curves in Fig.~2.b in the main article. The solid line is a fit to the high field part with critical exponent $\beta=0.368$ for 3D Heisenberg spins, as previously observed\cite{Wilhelm2016PRB_FeGescaling}. T$_N$ extrapolates to $278.2\,$K, also in-line with the literature\cite{Wilhelm2016PRB_FeGescaling}. \textit{(Pink triangles)} low field change in slope in $\Delta\rho_{yx}(H)$ curves, coinciding with the helical to conical transition. \textit{(Yellow triangles and circles)} left onset of the local minimum in $\Delta\rho_{yx}(H)$ and $\Delta\rho_{yx}(T)$, respectively. \textit{(White triangles and circles)} point of inversion between the minimum and maximum. \textit{(Red triangles and circles)} right onset of the maximum. \textit{(Dashed edge polygon)} Skyrmion lattice phase measured by SANS\cite{Moskvin2013PRLSANSFeGe} for H//[100] in a spherical bulk crystal after correcting for demagnetizing effects. From this, we attribute the maximum in $\Delta\rho_{yx}$ to the THE of the skyrmion lattice. No SANS data in the longitudinal geometry showing the six-fold scattering, is published at temperature below the open end of the polygon. The origin of the local minimum is still unidentified but it appears to continue into the inhomogeneous chiral spin state\cite{Wilhelm2016PRB_FeGescaling} between $T_N$ and the helical state. Dashed lines are guides to the eye.  

\section{Hysteresis cycle}
\begin{figure}
  \includegraphics[width=0.5\textwidth]{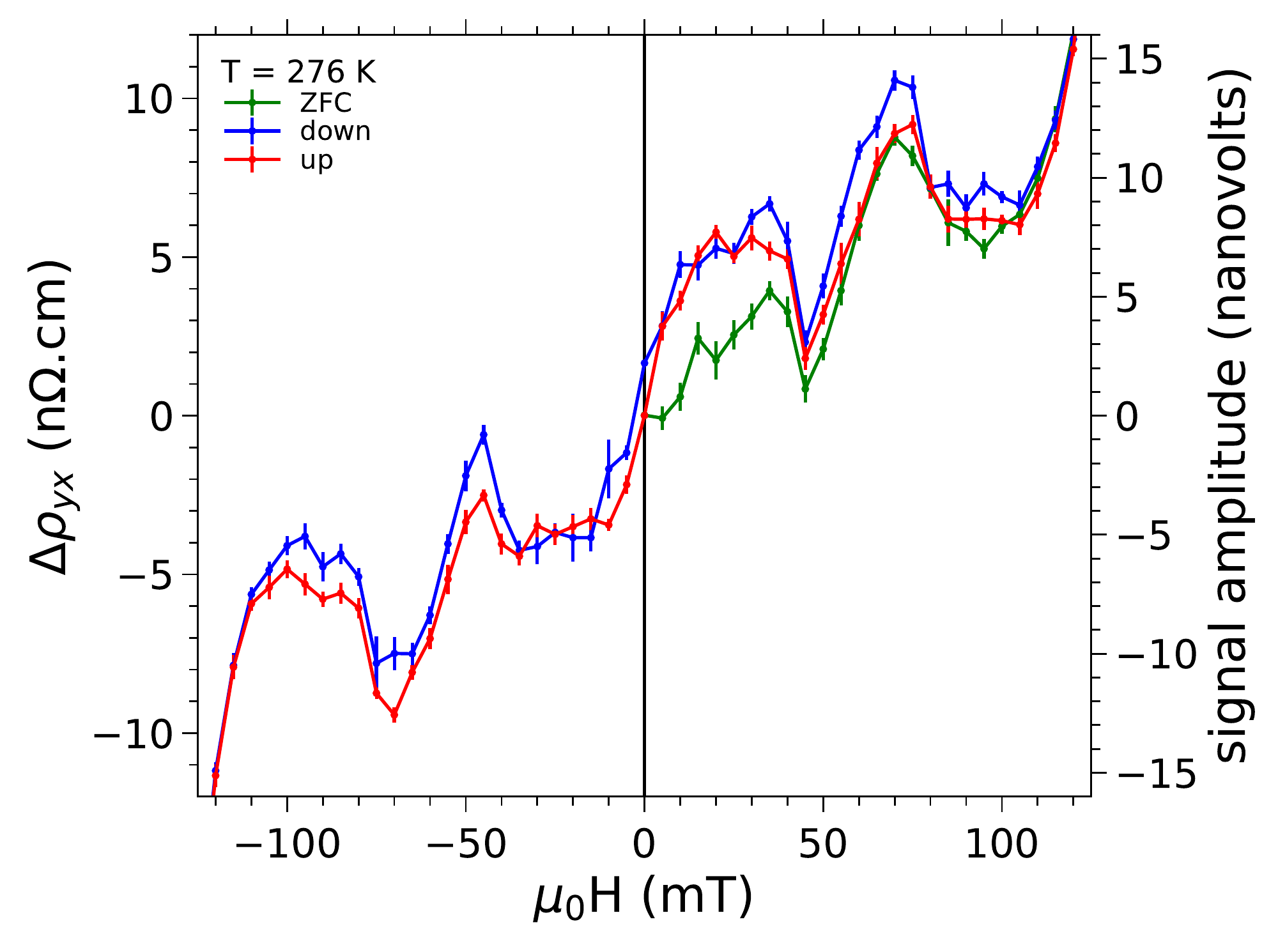}[h]
  \caption{\textbf{Deviation $\Delta\rho_{yx}(H)$ at 276\,K for a complete hysteresis cycle.}
}
  \label{fig3}
\end{figure}

A complete hysteresis cycle of $\rho_{yx}(H)$ was measured at 276\,K. We subtracted the slope $-4.3164\,$n$\Omega$.cm/mT from the $\rho_{yx}(H)$ curves measured during the hysteresis cycle, following the procedure explained in the main text (this slope corresponds to the 270-271\,K average slope of $\rho_{yx}(H)$ in the conical state). The resulting curves are shown in Fig.~\ref{fig3}. Compared to the $\rho_{yx}(H)$ curves, the only change in the initial ZFC ramp is an increase (from $\approx15$ to $\approx30$\,mT) in the upper bound of the helical phase. An extension of the helical phase in magnetic field, has also been observed in ZFC magnetization measurements in Fe and Co doped MnSi\cite{Bauer2010PRBMnFeSiMnCoSi}. 

\section{Demagnetizing factor}
To calculate the demagnetizing factors we follow the procedure detailed in Ref.~\citenum{Ritz2010PhDThesis}. The internal field is defined as
\begin{equation}
B_{int} = B_{ext} - \mu_0 N M
\end{equation}
where $B_{ext} = \mu_0 H$, $N$ is the demagnetizing factor and $M$ is the magnetization. In the conical state of FeGe, $M$ is linear with $H$ to a good approximation\cite{Wilhelm2016PRB_FeGescaling,Wilhelm2011PRLFeGemultiphase}. Stated otherwise, the magnetic susceptibility $\chi$, defined as $M = \chi H$, is a constant, hence
\begin{equation}
B_{int} = \mu_0 H (1- N \chi)
\end{equation}

But $\chi$ is geometry dependent. After determining $\chi^{sph}$ from magnetization measurement on a sphere, $\chi^{L}$ for the lamella sample can be deduced from Eqn.~(6.39) in Ref.~\citenum{Ritz2010PhDThesis}:
\begin{equation}
\chi^{L} = \frac{\chi^{sph}}{1-\chi^{sph}(N_{sph}-N_{L})}
\end{equation}
where $N_{sph}$ and $N_{L}$ are, respectively, the demagnetizing factors of the sphere and the lamella sample. 

To determine $\chi^{sph}$, we use the magnetization value $M = 0.20 \mu_B/$Fe for $\mu_0H = 0.05\,\mathrm{T}$ measured in a spherical sample and reported in Fig.~4 of Ref.~\citenum{Wilhelm2016PRB_FeGescaling} (M is constant in the conical state, in the measured range of temperatures from 265 to 280\,K). 
We also use: the unit cell volume from Ref.~\citenum{StoltJin2016nanolet_FeGeNW}, $V_{u.c.} = 103.82\ 10^{-30}$\,m$^{3}$, the fact that there are 4 formula units of FeGe per unit cell, and $\mu_B = 9.274\ 10^{-24}$\,J/T.
This yields
\begin{equation}
\chi^{sph} = \frac{dM}{dH} = \frac{\frac{0.20 *\mu_B}{\frac{V_{u.c.}}{4}}}{\frac{0.05}{\mu_0}} \approx 1.80\ (\mathrm{SI\ unit})
\end{equation}

Note that the value 1.80 for $\chi^{sph}$ that we deduced here from DC magnetization measurements is also in excellent agreement with the AC susceptibility measurements of Ref.~\cite{Wilhelm2011PRLFeGemultiphase} $\chi_{AC}=22\,$emu/mol for $\mu_0H_{AC}=1\,$mT, yielding $\chi^{sph}_{AC}=1.77$ in SI unit.

For a spherical FeGe sample such as the ones used in SANS\cite{Moskvin2013PRLSANSFeGe} and magnetization measurements\cite{Wilhelm2011PRLFeGemultiphase,Wilhelm2016PRB_FeGescaling}, $N_{sph}=1/3$ (see for instance Ref.~\cite{PhysRev.67.351} but note that it is in cgs units so that it finds $N = 4\pi\times1/3$, which is 1/3 in SI units).
Hence for the spherical sample:
\begin{equation}
B_{int} = \mu_0 H (1- N_{sph} \chi^{sph}) = \mu_0 H \left(1- \frac{1.8}{3}\right) = 0.4\, \mu_0 H
\end{equation}

For our lamella sample, $N$ is calculated from the dimensions of the sample using the analytical formula in Ref.~\cite{Aharoni1998a} for a ferromagnetic prism with arbitrary dimensions $a$, $b$ and $c$ : 
\begin{multline}
N = \frac{1}{\pi} \Bigg(\frac{a}{2 c} \log{\left (\frac{b + \sqrt{a^{2} + b^{2}}}{- b + \sqrt{a^{2} + b^{2}}} \right )} + \frac{b}{2 c} \log{\left (\frac{a + \sqrt{a^{2} + b^{2}}}{- a + \sqrt{a^{2} + b^{2}}} \right )}
 \\ + 2 \operatorname{atan}{\left (\frac{a b}{c \sqrt{a^{2} + b^{2} + c^{2}}} \right )} + \frac{c}{2 b} \log{\left (\frac{- a + \sqrt{a^{2} + c^{2}}}{a + \sqrt{a^{2} + c^{2}}} \right )} 
 \\ + \frac{1}{2 b c} \left(b^{2} - c^{2}\right) \log{\left (\frac{- a + \sqrt{a^{2} + b^{2} + c^{2}}}{a + \sqrt{a^{2} + b^{2} + c^{2}}} \right )} + \frac{c}{2 a} \log{\left (\frac{- b + \sqrt{b^{2} + c^{2}}}{b + \sqrt{b^{2} + c^{2}}} \right )}
 \\ + \frac{1}{2 a c} \left(a^{2} - c^{2}\right) \log{\left (\frac{- b + \sqrt{a^{2} + b^{2} + c^{2}}}{b + \sqrt{a^{2} + b^{2} + c^{2}}} \right )} + \frac{c}{a b} \left(\sqrt{a^{2} + c^{2}} + \sqrt{b^{2} + c^{2}}\right)
 \\ + \frac{\sqrt{a^{2} + b^{2} + c^{2}}}{3 a b c} \left(a^{2} + b^{2} - 2 c^{2}\right) + \frac{a^{3} + b^{3} - 2 c^{3}}{3 a b c}
 \\  + \frac{1}{3 a b c} \left(- \left(a^{2} + b^{2}\right)^{1.5} - \left(a^{2} + c^{2}\right)^{1.5} - \left(b^{2} + c^{2}\right)^{1.5}\right) \Bigg)
 \end{multline}
Using the dimensions $10\times25\times0.75\,\mu$m$^3$, we find $N_{L} \approx 0.8762$, so that

\begin{equation}
B_{int} = \mu_0 H \left(1- N_{L}\chi^{L})\right) = \mu_0 H \left(1- N_{L}\frac{\chi^{sph}}{1-\chi^{sph} (N_{sph}-N_{L})}\right) \approx 0.2023\,\mu_0 H
\end{equation}

Hence, for the lamella, an applied magnetic field of 70\,mT corresponds to $70\times0.2023\approx 14.1\,$mT of internal field, which corresponds to $14.1/0.4\approx 35.3\,$mT of applied magnetic field for the spherical sample measured in SANS.

\newpage